\documentclass{article}

\usepackage{algpseudocode}%
\usepackage[title]{appendix}%
\usepackage{authblk}
\usepackage{amsmath,amssymb,amsfonts}%
\usepackage{amsthm}%
\usepackage{arydshln}
\usepackage{booktabs}%
\usepackage{breakurl}
\usepackage{caption} 
\usepackage{float}
\usepackage{graphicx}%
\usepackage{latexsym}
\usepackage{listings}%
\usepackage{mathrsfs}%
\usepackage{manyfoot}%
\usepackage{multirow}%
\usepackage[round]{natbib}
\usepackage[caption = false]{subfig}
\usepackage{textcomp}%
\usepackage{url}
\usepackage{xcolor}%
\usepackage{xcolor}

\begin{document}

\author[1]{Leonardo Cefalo  } 
\author[1]{Crescenza Calculli }
\author[1]{Alessio Pollice }

\affil [1]{Department of Economics and Finance, Università degli Studi di Bari Aldo Moro, Bari, Italy}

\title{Modelling Access to Violence Against Women Support Centres in Apulia Using PC-Priors for CAR-Based Spatio-Temporal Models}

\maketitle

\begin{abstract}

Anti-violence centres are local non-governmental organisations designated to support women who suffered gender-based violence. In this study, we address the challenge of modelling the spatial variability in access to anti-violence centres across the municipalities of the Italian region of Apulia over the years 2021--2024 through Bayesian spatio-temporal Poisson regression. We investigate the association of municipality-level socioeconomic characteristics and local vulnerabilities with the frequency of access to the centres, and to explicitly account for the spatial dependence of the data-generating process, we compare three conditionally autoregressive latent spatial models -- intrinsic, proper and Leroux -- embedded in a spatio-temporal structure estimated via the Integrated Nested Laplace Approximation. Within this Bayesian hierarchical modelling framework, we propose an interpretable and principled hyperprior setting by assigning a joint penalised complexity prior to model parameters. Our methodological contribution is therefore the analytical derivation of the penalised complexity prior for the spatial association parameters in the proper and Leroux models, for which such priors were missing in the literature. We assess the relative fit of the competing models and prior specifications, discussing spatial confounding effects and inferential implications. In our case, spatial dependence is found to be moderate and the spatial association parameter retains appreciable posterior uncertainty, so that an interpretable and elicitable prior specification remains relevant. Lastly, our findings indicate that access to support services decreases with distance from the residential municipality, highlighting spatial constraints in help-seeking and the strategic importance of support centre location. Lower education levels are associated with lower access rates, which may suggest a potential under-reporting issue in disadvantaged areas, while higher economic development may be associated with a lower occurrence of violence, although the available data do not allow the two mechanisms to be disentangled.    

\end{abstract}

\section{Introduction}\label{sec.1}

Violence against women (VAW) is a widespread violation of human rights and a form of gender-based discrimination, encompassing acts causing physical, sexual, psychological, or economic harm. As recognised by the Istanbul Convention \citep{IstaCon2011}, this is a structural issue deeply tied to persistent gender inequalities worldwide, which makes accurate data essential to understand its scale and to develop effective policies. Despite this need, tracking gender-based violence remains challenging. In Italy, legislative initiatives such as Law No. 53/2022 \citep{GU2022} underscore the importance of systematic data collection for prevention, yet data sources remain fragmented, often outdated, and not easily accessible. Stigma and under-reporting further hinder accurate quantification, as many survivors fear social repercussions, blame, or disbelief. 

Local anti-violence centres (AVCs), established under Italian Law No. 119/2013 \citep{GU2013}, play a crucial role in supporting victims through a holistic array of services, starting from consultancy and including psychological support, legal assistance, and sheltering, while also documenting cases of abuse.
In 2017, the Italian National Institute of Statistics (ISTAT) started monitoring the activity of AVCs \citep{ISTAT2023}, capturing the scope and characteristics of violence experienced by women seeking help across the national territory. From then until 2023, the number of AVCs active over the national territory increased by 43.8\%, with 404 AVCs operating in 2023. Similarly, the number of women accessing them increased by 41.5\%. This information may suggest that the expansion of AVC activity encourages women victims of violence to seek help and find a way out of violence and abuse.

Records from AVCs provide valuable information on victim demographics and violence patterns \citep{Toffanin2020}, enabling investigation of the distribution of the phenomenon at the territorial level and allowing us to assess how local socioeconomic factors affect the usage of support services, which reflects both the occurrence of violence and the propensity to seek help.

The process of finding a way out of violence may indeed start with seeking help from AVCs \citep{CAV2024report} and, more generally, from sheltering structures, hence shelter accessibility is a crucial requirement. The isolation of peripheral areas and the resulting lower accessibility to shelters may hinder help-seeking, as noticed by \cite{peek2011rural}. In addition, \cite{Denti2022} found strong evidence that local support services in the United Kingdom have a positive impact on gender-based violence reporting, even when controlling for the implementation of nationwide measures. Similarly, in Italy, \cite{Bettin} examined the \textit{Codice Rosa} initiative (a triage code dedicated to hospitalising VAW victims) and anti-stalking help desks, highlighting the significant impact of the former on the reporting of violence nationwide.

Building on this evidence, we investigate how local disparities in the accessibility of AVCs may shape women’s capacity to seek help and escape situations of violence by studying the records of access to centres within the Apulia region (Southern Italy), provided by the regional authority itself and covering the years 2021--2024. Specifically, we propose a territorial analysis of accesses detailed by the origin municipality of women seeking help, using a Bayesian hierarchical spatio-temporal Poisson regression model \citep{martinez2008autoregressive}. Our main goal is to investigate how accesses are associated with both the distance from AVCs and help desks -- measured through the road travel time between origin municipalities and municipalities hosting either AVCs or help desks -- and a set of socioeconomic indicators employed to define the Municipality Fragility Index provided by the ISTAT \footnote{\label{note:MFI} Available on the ISTAT website \url{https://www.istat.it/comunicato-stampa/aggiornato-indice-di-fragilita-comunale/}}.

\begin{figure}[ht]
\centering
\subfloat{\includegraphics[width = 2.5in]{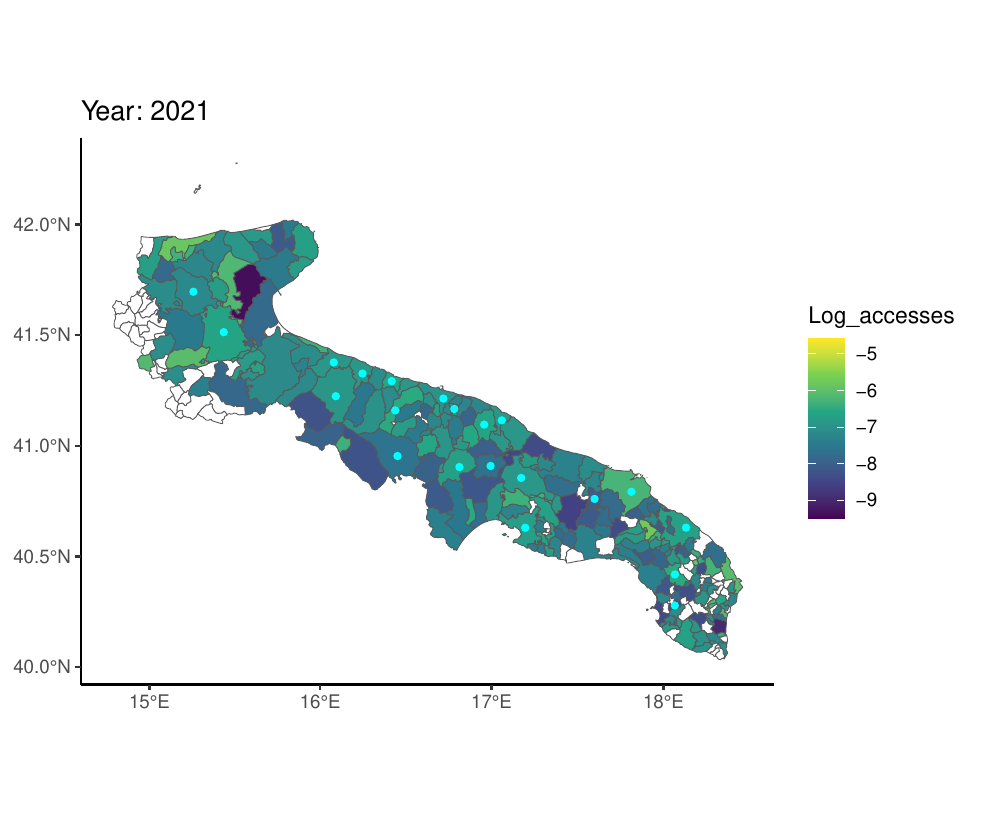}} 
\subfloat{\includegraphics[width = 2.5in]{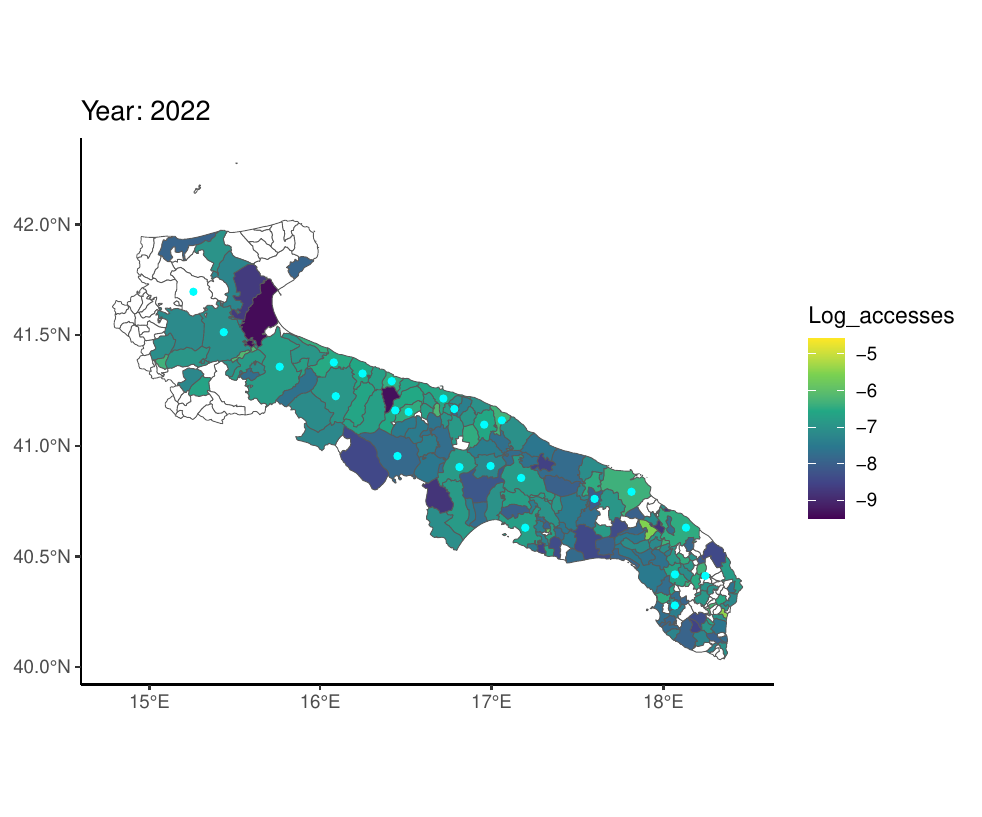}}\\
\subfloat{\includegraphics[width = 2.5in]{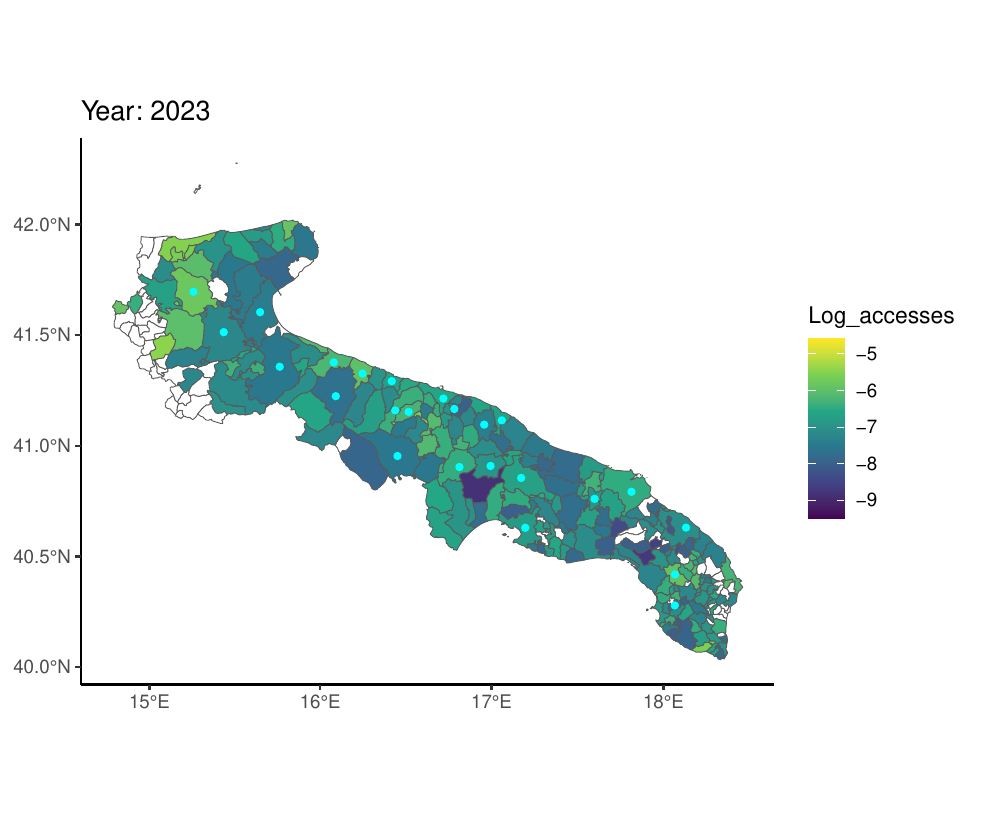}}
\subfloat{\includegraphics[width = 2.5in]{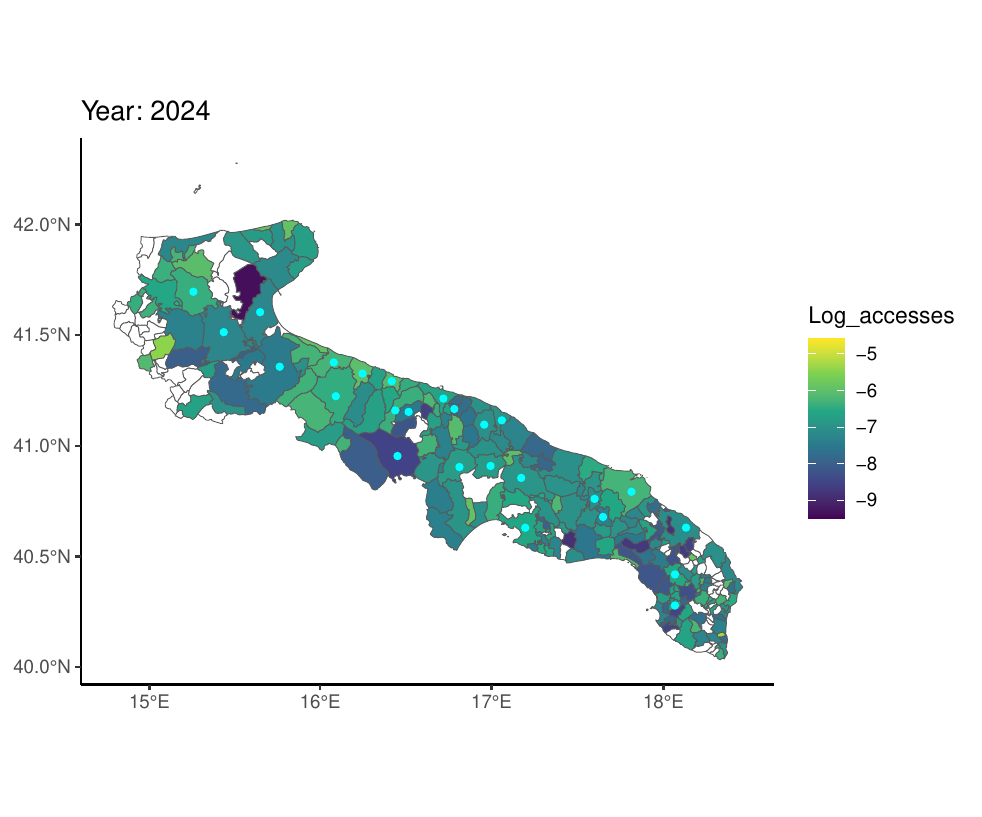}} 
\caption{Log-frequency of AVC accesses. White areas correspond to zero counts. Points correspond to municipalities hosting at least one AVC.}\label{fig:log_access}	
\end{figure}

Figure~\ref{fig:log_access} shows the frequency of AVC accesses, defined as the count of accesses over the female population aged $>14$ years, across Apulian municipalities in the 4 years on the logarithmic scale, alongside the municipalities with at least one operating AVC. White areas correspond to zero counts. This map suggests a heterogeneous spatial pattern of access, with slightly higher values for coastal municipalities, indicating potential differences in the population, infrastructure, and service availability between coastal and inland areas. 

The inherent spatial dependence in the data-generating process suggests that we introduce a latent spatial process in the regression model. We model this latent process as a Gaussian Markov random field defined on the municipality adjacency structure, and consider three alternative spatial structures: the Intrinsic Conditional Autoregressive model \citep[ICAR,][]{Besag1995}, the Proper Conditional Autoregressive model \citep[PCAR,][]{Gelfand2003}, the Leroux Conditional Autoregressive model \citep[LCAR,][]{Leroux1999}. Moreover, as observations cover multiple years, we integrate these spatial structures into spatio-temporal models \citep{martinez2008autoregressive}.

The spatio-temporal models depend in turn on an additional layer of hyperparameters, which require some prior distributional assumptions. In this work, we propose a flexible and interpretable prior setting based on the principle of parsimony, which can be fulfilled by penalising \textit{a priori} the departure of these flexible models from simpler ones lacking either the spatial structure, the temporal structure, or the presence itself of a latent component. This can be done intuitively through the Penalised Complexity prior (PC-prior) approach, developed by \cite{PC}. 

To the best of our knowledge, the PC-prior approach has not yet been employed for the hyperparameters controlling the strength of spatial association in the PCAR and LCAR models. However, we find that the PC-prior can be derived analytically for these models, which enables us to propose a full PC-prior specification and compare it with the case in which the spatial association parameter is assigned a Uniform prior.
To enable a proper operational comparison of the proposed priors, we consider the class of CAR-type models and exclude the Besag-York-Mollié (BYM) model \citep{Besag1991, BYM2} from our analysis, as it features a slightly different formulation from the CAR-type models, which are based on a Markovian structure and feature a parameter measuring spatial association, while the BYM is based on the convolution of a spatially structured and an IID component. Moreover, the PC-prior setup of the BYM model is currently well-known and has been developed in the seminal papers of  \cite{PC} and \cite{BYM2}.

Spatial confounding is another relevant issue encountered with the analysis of the Apulia data, as explicitly accounting for unmeasured spatial variability caused some potential confounding bias in the estimation of the effects of some explanatory variables. This is known to happen when the explanatory variables show a spatial trend \citep{RHZ}. For those covariates whose effects differ most markedly between the nonspatial and the corresponding spatial regression, we mitigate this bias using a simplified version of the Spatial+ method \citep{Dupont} that captures the spatial trends through the lowest-frequency eigenvectors of the graph Laplacian matrix \citep{Urdangarin24}.

Throughout this paper, model estimation is carried out within the Integrated Nested Laplace Approximation \citep[INLA,][]{INLA, VBINLA2023}, which enables computationally efficient Bayesian inference and model selection based on several selection criteria, among which we choose to use the Watanabe-Akaike Information Criterion \citep[WAIC,][]{GelmanWAIC}. Considering the present data and the models in scope, we find that the model leading to the smallest WAIC is the Leroux CAR, and that the PC-prior slightly outperforms the more familiar Uniform prior; this also holds when applying the simplified Spatial+ correction to explanatory variables.

The proposed inferential framework and the available data allow us to identify differences in observed access rates and discuss whether the current distribution of services may leave some areas at a disadvantage. Specifically, the study region is characterised by significant socioeconomic disparities \citep{PROMETHEEII}, including heterogeneity in economic development, infrastructure, and service accessibility. This circumstance, together with the high level of territorial detail in available data provided by the Regional Statistical Office, makes Apulia an ideal case study to investigate how such factors affect the search for AVC assistance. 

The contribution we propose in this work is therefore twofold. On the methodological side, we present a principled and interpretable prior for the spatial association parameters of the PCAR and LCAR models. These priors are found to depend only on the spatial dependency structure through the eigenvalues of two deterministic matrices, in analogy with the variance mixing parameter in the scaled Besag-York-Molli\'{e} model \citep{BYM2, PC}. Moreover, we notice that the PC-prior for the AR(1) autocorrelation does not depend on the spatial structure but only on the number of time points, as proved by \cite{Sorbye2017} for the purely temporal case. Therefore, the hyperpriors we employ are directly transferable to other CAR-based spatio-temporal applications. On the application side, we provide a territorial analysis of access to anti-violence centres, quantifying the association with centre accessibility and local socioeconomic vulnerability while accounting for unmeasured spatial effects and potential spatial confounding.

The modelling strategy can be summarised as follows:\\
\textit{i)} A spatio-temporal Poisson regression is specified for municipality-level access counts, with a linear predictor combining covariate effects and a latent spatiotemporal field. \\
\textit{ii)} The latent field is modelled using three alternative CAR specifications, ICAR, PCAR, and LCAR, which we embed in turn in an AR(1) temporal structure; for comparison, we also estimate a plain Poisson model with no latent field (nonspatial). \\
\textit{iii)} PC-priors are defined on the temporal autocorrelation and scale parameters of the latent fields; for the spatial association parameter in the PCAR and LCAR models we compare three alternative priors, i.e. the Uniform prior and two PC-prior specifications. \\
\textit{iv)} Each spatio-temporal specification is estimated either retaining the
explanatory variables as they are, or filtering out the 15, 20 or 25 smoothest Laplacian
eigenvectors from the covariates affected by spatial confounding. \\
\textit{v)} The resulting model configurations are compared through the WAIC. 

The remainder of this paper is structured as follows:  section \ref{section:models} describes the Bayesian spatiotemporal modelling framework and provides some generalities on the proposed PC-priors, section \ref{section:data} introduces the data set of accesses to AVCs in Apulia and relevant explanatory variables, section \ref{sec:imple} addresses some practical implementation issues for the data at hand and summarises the strategy to reduce spatial confounding, and section \ref{sec:results} provides some results of posterior inferences and interpretative comments.

\section{Spatiotemporal Poisson modelling framework}\label{section:models}

Throughout this study, we consider a spatio-temporal Poisson model that for site $i \in \lbrace1, \dots, n\rbrace $ and time period $t \in \lbrace1, \dots, T\rbrace$ takes the following form:
$$
y_{i, t} \sim \mathrm{Poisson}(P_{i,t} \,e^{ \eta_{i,t} })
$$
where $P_{i,t}$ is an offset term and $\eta_{i,t} := \beta_{0, t} +\boldsymbol{x_{i,t}^{\top}\beta} + z_{i,t}$ is a linear predictor; $\boldsymbol{x_{i,t}}$ is a vector of known explanatory variables, $\boldsymbol{\beta}$ is a vector of constant covariate effects, $\beta_{0,t}$ is a time-dependent intercept, and $z_{i,t}$ is a realisation of a latent spatio-temporal field $\boldsymbol{z} = (\boldsymbol{z_{1}^\top}, \dots, \boldsymbol{z_{T}^\top})^{\top}$. Following the terminology of \cite{KnorrHeld}, four types of interactions can be distinguished for spatio-temporal random fields: temporally and spatially independent (type I); temporally structured and spatially independent (type II); temporally independent and spatially structured (type III) and both temporally and spatially structured (type IV). Here, $\boldsymbol{z}$ is modelled as a type IV interaction. Specifically, following \cite{martinez2008autoregressive} we assume a lag-1 temporal dependence and model $\boldsymbol{z}$ as:
\begin{equation}\label{eq:AR1}
\left \{ \begin{array}{ll}
 \boldsymbol{z_t} = r \boldsymbol{z_{t-1}} + \boldsymbol{\nu_t}  \quad \text{for} \quad t=2,\dots, T\\
\boldsymbol{\nu_t} \sim \mathcal{N}_n\left(0, \sigma^2 \boldsymbol{\Omega}\right)  \quad \text{for} \quad t=2,\dots, T\\
\boldsymbol{z_1} \sim \mathcal{N}_n \left( 0, \frac{1}{1-r^2}\sigma^2 \boldsymbol{\Omega} \right)
\end{array} 
\right.
\end{equation}
where $\boldsymbol{\nu_t}$ is a spatially structured Gaussian random field taking values over a graph of order $n$, described by the symmetric and positive semi-definite $n\times n$ spatial variance-covariance matrix $\boldsymbol{\Omega}$, $\sigma^2$ is a marginal variance parameter, and $r$ is a temporal correlation parameter. In this work, the nodes of the reference graph are the $n=256$ Apulian mainland municipalities.

Assuming that $\boldsymbol{\Omega}$ and $\sigma^2$ are constant over time, the joint distribution of $\boldsymbol{z}$ reads as
$$
\boldsymbol{z} \sim \mathcal{N}_{T n} \left(0, \frac{\sigma^2}{1-r^2}\boldsymbol{R} \otimes \boldsymbol{\Omega }\right)
$$
where $\boldsymbol{R}$ is the AR(1) correlation matrix of size $T \times T$ with elements $R_{st}=r^{\lvert s-t \rvert}$ for $s, t \in \lbrace1, \dots, T\rbrace$. This choice is justified by the fact that the present formulation constitutes a simplification of a more general multivariate spatial model in which $T$ marginal variances and $T(T-1)/2$ correlations would be required. The $\otimes$ operator denotes the Kronecker product.
Moreover, since we have no prior information justifying a temporal variation in the scale or spatial association parameters, we assume the distribution of $\boldsymbol{\nu_t}$ in Equation (\ref{eq:AR1}) to be stationary over time. 
Besides the spatial association parameter, the proposed spatio-temporal model thus employs only one scale parameter $\sigma^2$ and one correlation parameter $r$, resulting in a simplified between-years variance-covariance matrix and a sparse between-years precision.

\subsection{Spatial structures} \label{subsection:spatial}

The regression model outlined in the previous section features a spatial component described by the matrix $\boldsymbol{\Omega}$. In the following, we outline some features of the three CAR-type spatial models used in the analysis.

\subsubsection{Intrinsic Conditional Auto Regressive}  \label{par:ICAR}
Considering a graph described by a binary adjacency matrix $\boldsymbol{W}$, with $w_{ij} = 1$ if areas $i$ and $j$ share a border and $w_{ij}=0$ otherwise, with $i,j \in \lbrace1, \dots, n \rbrace$, at time $t \in \lbrace1 , \dots,T\rbrace$ the variable $\nu_{i,t}$ has the conditional distribution:
$$
\nu_{i,t}\vert \boldsymbol{\nu_{-i, t}},\sigma \sim 
\mathcal{N} \left(\sum_{j = 1}^n \frac{w_{i,j}}{d_i}\nu_{j,t},
\, \frac{\sigma^2}{d_i} \right)
$$
where $\boldsymbol{\nu_{-i, t}}:= \left(\nu_{1,t}, \dots, \nu_{i-1, t}, \nu_{i+1,t}, \dots, \nu_{n,t} \right)$, while $d_i:=\sum_{j=1}^n w_{ij}$ denotes the number of neighbours of the $i$-th site. Lastly, $\sigma$ is a parameter controlling the scale of the spatio-temporal process. For this model the joint spatial precision matrix coincides with the $n \times n$ graph Laplacian matrix $\boldsymbol{L} = \boldsymbol{D} - \boldsymbol{W}$, where $\boldsymbol{D}:= \mathrm{diag}\left(d_1, \dots, d_n \right)$ is the diagonal graph degree matrix. Since the Laplacian matrix of the ICAR is always singular, the spatial covariance matrix is given by the Moore-Penrose pseudoinverse: $\boldsymbol{\Omega} := \boldsymbol{L^+}$. As noted, for example, by \cite{Besag1995}, the ICAR model is not identifiable unless a linear constraint is imposed; as a consequence, $\boldsymbol{\nu_t}$ and $\boldsymbol{z_t}$ are constrained to sum to zero for each $t$.

\subsubsection{Proper Conditional Auto Regressive} 
This structure extends the flexibility of the ICAR by introducing the spatial autocorrelation parameter $\rho \in [0,1]$ to control the strength of spatial association. In this case, the conditional distribution of the generic $\nu_{i,t}$ element is given by:
$$
\nu_{i,t}\vert  \boldsymbol{\nu_{-i,t}}, \sigma, \rho \sim 
\mathcal{N} \left(\sum_{j = 1}^n \frac{\rho w_{i,j}}{d_i}\nu_{j,t},
\, \frac{\sigma^2}{d_i} \right)
$$
Then, the joint prior of $\boldsymbol{\nu_t}$ has spatial covariance matrix given by
\begin{equation}\label{eq:PCAR}
\boldsymbol{\Omega} := \left(\boldsymbol{D} - \rho \boldsymbol{W}\right)^{-1}
\end{equation}
Clearly, the limit case for $\rho = 1$ is the ICAR, while $\rho = 0$ implies an independent but not identically distributed Gaussian random field. Even though this model does not require the sum-zero identifiability constraint, in our analysis we apply it in analogy with other models to make results comparable. 
\\

\subsubsection{Leroux proper Conditional Auto Regressive} 
This model can be considered as a different extension of the ICAR, introducing the spatial association parameter $\lambda \in [0, 1]$. Here, the conditional prior takes the form
$$
\nu_{i,t} \vert \boldsymbol{\nu_{-i,t}}, \sigma, \lambda \sim
\mathcal{N} \left(\sum_{j=1}^{n} \frac{ \lambda w_{i,j}}{1 - \lambda  + \lambda d_i} \nu_{j,t},\,
\frac{\sigma^2}{1 - \lambda + \lambda d_i} \right)
$$
and the resulting joint spatial covariance matrix is given by
\begin{equation}\label{eq:LCAR}
\boldsymbol{\Omega} := \left(\lambda \boldsymbol{L} + (1-\lambda) \boldsymbol{I_n} \right)^{-1}
\end{equation}
The limit case for $\lambda = 1$ is again the ICAR, while for $\lambda = 0$ the model reduces to an IID Normal field. While this is not a singular model, we follow \cite{Goicoa} and apply the sum-zero constraint to ensure that the model is identifiable and that no confounding occurs with the intercept.

\subsection{The PC-prior framework} \label{subsec:PC}

The regression model outlined so far features a latent spatio-temporal effect, which implies a potentially high degree of complexity. In the context of Bayesian hierarchical models, a natural approach to measuring and penalising model complexity is provided by PC-priors \citep{PC}. 

The basic idea behind the PC-prior for a hyperparameter $\theta$ is to penalise \textit{a priori} the departure of a flexible model $\mathcal{M}_1$ with random $\theta$ against a base model $\mathcal{M}_0$ in which $\theta$ is fixed at a given value, say $\mathcal{M_0}: \theta = \theta_0$. \cite{PC} propose to measure the departure between the two models with the distance function $\sqrt{2 \mathrm{KLD}(\mathcal{M}_1 || \mathcal{M}_0)}$, where $\mathrm{KLD}\left(\cdot||\cdot \right)$ denotes the Kullback-Leibler divergence. In our case, hyperparameters affect the prior distribution of $\boldsymbol{z}$ and the models to be compared are $\mathcal{M}_1:\boldsymbol{z}\sim \mathcal{N}\left( \boldsymbol{0}, \boldsymbol{\mathrm{Var}_1} \right)$ and  $\mathcal{M}_0:\boldsymbol{z}\sim \mathcal{N}\left( \boldsymbol{0}, \boldsymbol{\mathrm{Var}_0} \right)$ where the matrix $\boldsymbol{\mathrm{Var}_1}$ is a function of $\theta$ and $\boldsymbol{\mathrm{Var}_0}$ equals the same function evaluated at $\theta = \theta_0$. Therefore, the KLD for $\theta$ measures the divergence between two densities of $\boldsymbol{z}$, i.e. $\mathrm{KLD}\left(\mathcal{M}_1 \lvert \rvert \mathcal{M}_0 \right) = \int_{-\infty}^\infty \left[ \ln \pi(\boldsymbol{z} \vert \theta) - \ln \pi(\boldsymbol{z} \vert \theta = \theta_0) \right] \pi(\boldsymbol{z} \vert \theta) d \boldsymbol{z}$.

Additionally, \cite{PC} propose to penalise the distance between the two models by assigning an exponential prior to the distance function. By doing so, the exponential prior depends on a rate parameter, say $\psi$, that can be tuned to assume \textit{a priori} that a certain inequality between $\theta$ and a boundary value $U$ has a given probability of realisation $\alpha$ (e.g. $\theta \leq U$ with probability $\alpha$).

In this paper, we define a joint PC-prior setup for the whole array of hyperparameter. For the spatio-temporal ICAR model, hyperparameters are only the temporal autocorrelation $r$ and the standard deviation $\sigma$. The LCAR and PCAR models employ an additional hyperparameter that represents the strength of the spatial association, $\rho$ or $\lambda$, respectively. Since, to the best of our knowledge, the derivation of the PC-prior for the latter parameters constitutes a novelty, we also test the Uniform prior on such parameters as a more familiar alternative.

Overall, the joint prior on the whole hyperparameter set reads as:
$$
\pi(r, \sigma, \xi) =\pi(r \mid \sigma, \xi) \pi(\xi\mid\sigma) \pi(\sigma)
$$
where $\xi \in \lbrace \rho, \lambda \rbrace$. As discussed in more detail in Appendix \ref{Appendix:PCpriors}, the KLD of both $r$ and $\xi$ can be seen to be independent of any other parameter, implying that $r$, $\sigma$, and $\xi$ are independent \textit{a priori} and their joint prior can be factorised into the three relevant marginals. The PC-prior for $\sigma$ has been described by \cite{PC} and proved to coincide with an Exponential prior. In the following sections, we provide some details on how the priors are defined for the other hyperparameters.

\subsubsection{AR(1) autocorrelation parameter} \label{subsection:PC_ar1}

We obtain the PC-prior for the temporal autocorrelation parameter $r$ in the AR(1) model in (\ref{eq:AR1}) penalising $r \in [-1, 1]$ against $r=0$. Given a boundary value $U \in [0, 1]$ and a central probability $\alpha$, such that $\mathrm{Prob} \lbrace \lvert r \rvert\leq U \rbrace = \alpha$, the prior takes the form:
\begin{equation}
  \pi(r) = \ln \left(\frac{1}{ 1-\alpha}\right)
  \frac{1}{4\sqrt{h(U)h(r)}} \left(1-\alpha\right)^{
  \frac{\sqrt{ h(r)}}{ \sqrt{h(U)}} 
  }  \left| \frac{\partial h(r)}{\partial r} \right|
  \label{eq:PC_AR1}
\end{equation}
where the function $h(r) = \ln(1-r^2) + T  \frac{r^2}{1-r^2}$ does not depend on the graph structure or the number of sites $n$. This is indeed a desirable property, as it would make no sense for the temporal autocorrelation parameter $r$ to depend on $n$. This finding extends to the spatio-temporal case the  results of the purely temporal AR(1) autocorrelation parameter discovered and discussed by \cite{Sorbye2017}. The calculations leading to Equation \ref{eq:PC_AR1} are shown in Appendix \ref{Appendix:PCpriors}. 

\subsubsection{PCAR spatial association parameter} \label{subsection:PC_rho}

The KLD for the PCAR spatial association parameter $\rho$ in Equation \ref{eq:PCAR} is defined by comparing $\mathcal{M}_1 : \boldsymbol{\Omega} = \left(\boldsymbol{D}-\rho \boldsymbol{W}\right)^{-1}$, with $\rho \in [0,1]$ against $\mathcal{M}_0:\rho = 0$, i.e.  $\boldsymbol{\Omega} = \boldsymbol{D^{-1}}$. Assuming $\boldsymbol{D}$ is nonsingular, the KLD reads as:
  \begin{equation}
\mathrm{KLD}\left(\pi\left(\boldsymbol{z} \vert\rho \right)\, || \,\pi \left(\boldsymbol{z} \vert\rho=0\right)\right) = \frac{T}{2} 
\sum_{i=1}^n  \left(
\ln(1  - \rho \delta_i) + \frac{1}{1-\rho \delta_i} \right) - \frac{Tn}{2}
 \label{eq:KLD_PCAR}
 \end{equation}
where $\delta=(\delta_1,\ldots,\delta_n)$ is the array of eigenvalues of the deterministic matrix $\boldsymbol{D^{-1}W}$. More details on the calculation of the KLD and its derivative are in appendix \ref{appendix_pcar}. To our aims, it is worth noting that the KLD, and therefore the PC-prior, of $\rho$ is independent of other model parameters and requires a computational cost of order $n$.

\subsubsection{LCAR spatial association parameter} \label{subsection:PC_lambda}

In this case, we compare $\mathcal{M}_1 : \boldsymbol{\Omega} = (\lambda \boldsymbol{L} +(1-\lambda)\boldsymbol{I_n})^{-1}$, with $\lambda \in [0,1]$ against $\mathcal{M}_0:\lambda = 0$, i.e. $\boldsymbol{\Omega} = \boldsymbol{I_n}$. The KLD is given by:
  \begin{equation}
\mathrm{KLD}\left(\pi\left(\boldsymbol{z} \vert\lambda \right)\, || \,\pi \left(\boldsymbol{z} \vert\lambda=0\right)\right) = \frac{T}{2} 
\sum_{i=1}^n  \left(
\ln \left[ \lambda (\ell_i - 1) + 1 \right] + \frac{1}{\lambda (\ell_i - 1) + 1} \right) - \frac{Tn}{2}
 \label{eq:KLD_LCAR}
 \end{equation}
where $\ell=(\ell_1,\ldots,\ell_n)$ is the array of eigenvalues of the graph Laplacian matrix $\boldsymbol{L}$. In analogy with the PCAR, the KLD requires a linear computational cost as Laplacian eigenvalues are deterministic. More details are in appendix \ref{appendix_lcar}

\subsection{Spatial confounding treatment} \label{subsection:spatial+}

When the assumption of independence between explanatory variables and latent spatial effects is violated, the issue of spatial confounding may arise, causing bias in the estimates of covariate effects. In this work, we apply the simplified Spatial+ methodology developed by \cite{Urdangarin24} to mitigate spatial confounding in areal spatial data settings. 

Specifically, let us express a generic covariate $\boldsymbol{x}$ as $\boldsymbol{x} =\boldsymbol{V b}$, where $\boldsymbol{V}$ are the eigenvectors of $\boldsymbol{L}$ and $\boldsymbol{b}$ is an appropriately chosen vector of length $n$. With no loss of generality, eigenvectors are sorted in descending order of the associated eigenvalues, i.e. the last eigenvectors are the smoothest ones \citep{SpectralClust}. 
We partition the Laplacian eigenvector matrix as $\boldsymbol{V} =\left( \begin{array}{c:c} \boldsymbol{V^{(ns)}} &\boldsymbol{V^{(s)}}\end{array} \right)$ where the nonspatial component $\boldsymbol{V^{(ns)}}$ includes the first $n-k$ eigenvectors and the spatial component $\boldsymbol{V}^{(s)}$ includes the last $k$ eigenvectors for an integer $k$ strictly smaller than $n$; likewise, we partition the vector $\boldsymbol{b}$ into $\left( \begin{array}{c:c} {\boldsymbol{b^{(ns)}}}^\top & {\boldsymbol{b^{(s)}}}^\top \end{array}\right)^\top$. Then, we can decompose $x$ as the sum of the nonspatial component $\boldsymbol{x^{(ns)}}:=\boldsymbol{V^{(ns)} b^{(ns)}}$ and the spatial component $\boldsymbol{x^{(s)}}:=\boldsymbol{V^{(s)} b^{(s)}}$, and in the regression model in section \ref{section:models} we employ $\boldsymbol{x^{(ns)}}$ as the explanatory variable. To the best of our knowledge there is no rigid rule to define $k$; in this work, we test $k \in \lbrace 15, 20, 25 \rbrace$, i.e. we retain at least 90\% of Laplacian eigenvectors, considering $n=256$ areas, which correspond to mainland Apulian municipalities.

\section{The Apulia Region data}\label{section:data}

Data on women's access to AVCs and support services are collected by the Statistical Section and the Welfare Department of the Apulia Region as part of a systematic monitoring programme established under Regional Law No. 29/2014. Since 2013, all AVCs were required to submit annual reports detailing women’s access to support services to monitor the effectiveness of the service and identify local problems. The resulting data collection and processing activities contribute to the Regional annual reports \citep{CAV2024report}.

We focus on the number of accesses to AVCs in the $256$ mainland Apulian municipalities over the years 2021 -- 2024. To avoid considering duplicate data records, we only retain reports of violence that are directly managed by each centre, excluding those that are redirected to other services or centres. In Apulia, the total number of women supported by AVCs was 1477, 1516, 1822, 1778 over the four years in scope.

Women accessing AVCs may report up to three types of violence explicitly; specifically, only 22.5\% of them reported only one type of violence, while 41.4\% and 35.6\% reported two and three types of violence co-occurring, respectively. Psychological, physical and economic violence are the most frequently reported types, recorded for 83.3\%, 64.5\% and 28.1\% of women, respectively. In most violence records, the perpetrator is the woman's partner; specifically, spouses, ex-spouses, cohabiting partners and ex-cohabiting partners have been indicated as aggressors by 32.3\%, 14.8\%, 11.7\% and 11.1\% of women, respectively.

\begin{figure}[ht]
\centering
\subfloat{\includegraphics[width = 1.75in]{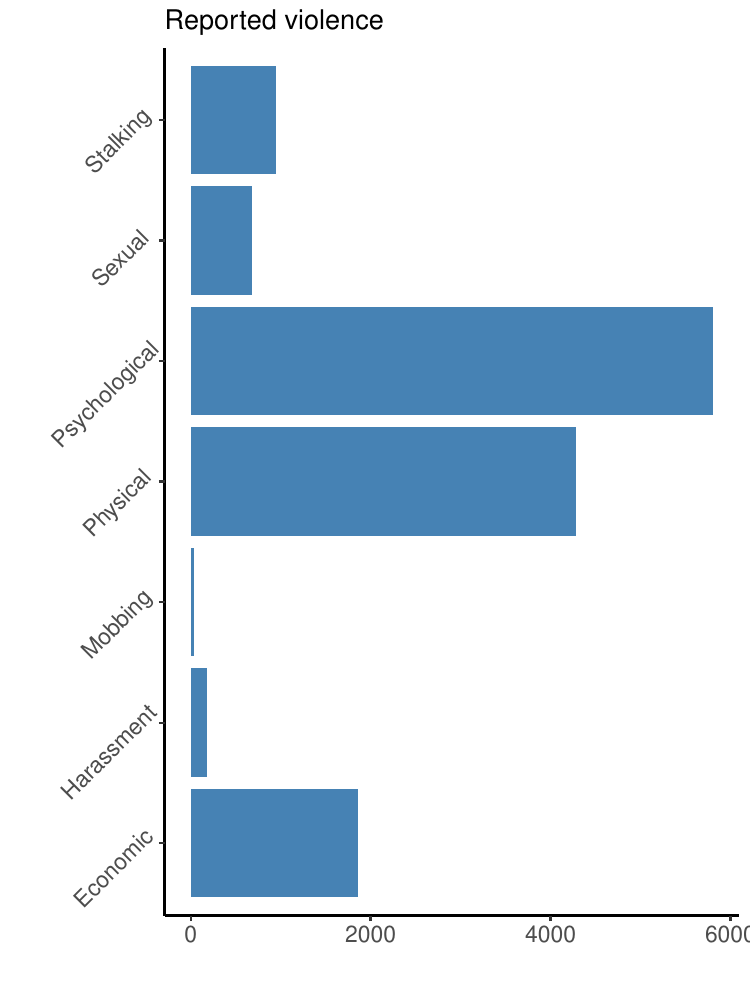}}
\subfloat{\includegraphics[width = 1.75in]{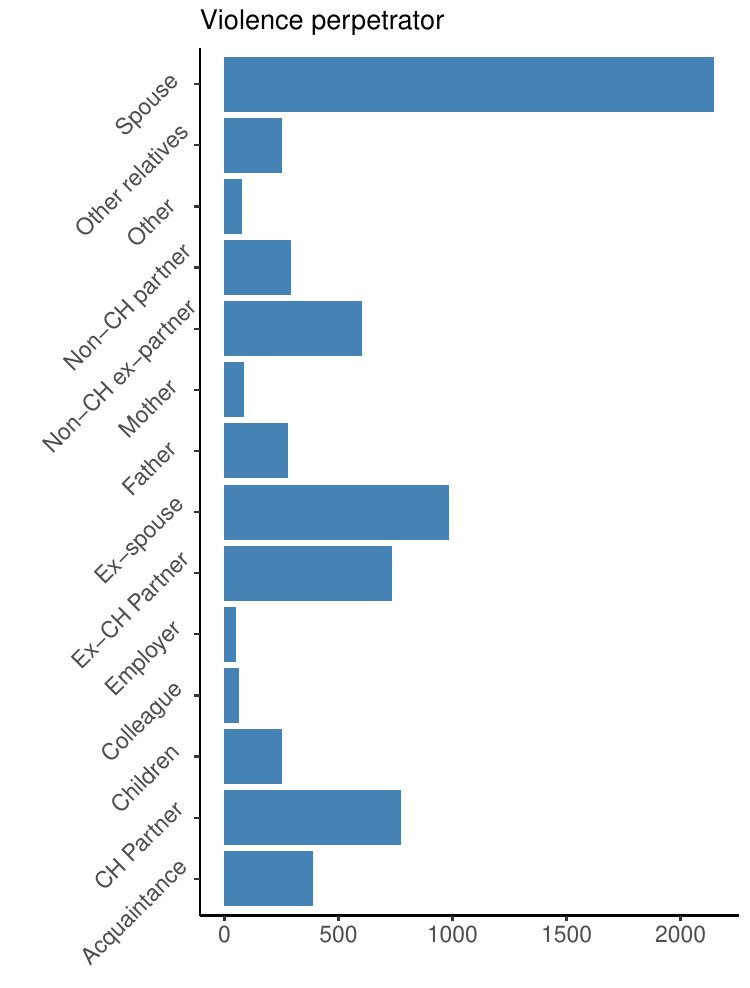}}
\subfloat{\includegraphics[width = 1.75in]{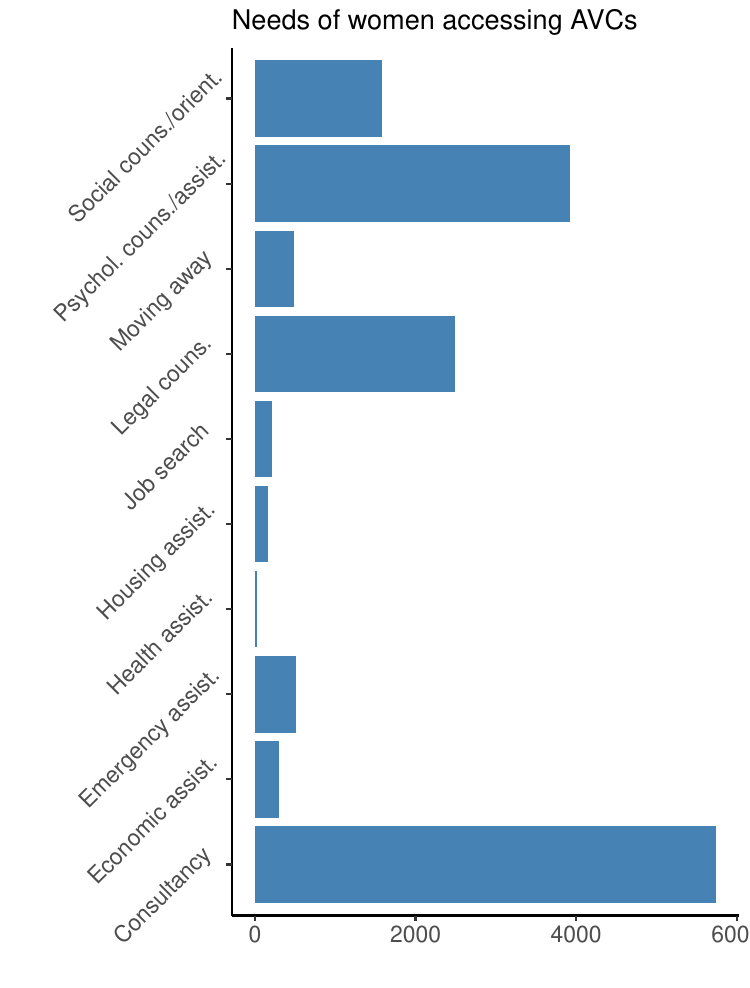}}
\caption{Counts of types of reported violence (left), relation of the aggressors to women seeking help (middle; "CH" stands for cohabiting), and types of help requested from AVCs (right) over the years in scope.}\label{fig:barplot}	
\end{figure}

Concretely, women accessing AVCs seek help through services, namely initial consultancy; economic, psychological, social, housing, legal assistance; job search; moving away from potentially dangerous partners. Besides initial consultancy, which is needed by almost all women (86.4\%), the most requested services are psychological counselling or assistance, legal assistance and social counselling or orientation (59\%, 37.5\% and 23.8\% of access records, respectively). In more detail, Figure \ref{fig:barplot} shows the counts of reported violence type, relationship to the aggressor and type of help requested over the years in scope. 

Subsequent analyses focus on modelling the spatial distribution and temporal evolution of the number of accesses to AVCs using the Poisson regression model defined in section \ref{section:models}, considering the female population aged $>14$ years\footnote{Data retrieved from ISTAT at: \url{http://dati.istat.it/Index.aspx?DataSetCode=DCIS_POPRES1}} as offset. 
To investigate the association of access rates with potential drivers at the municipal level, a set of candidate explanatory variables is considered. First, we employ the distance of women's residence municipalities from the closest municipality hosting either an AVC or a help desk; between-municipalities road distances are provided by ISTAT \citep{ISTATDIST}\footnote{Data retrieved from \url{https://www.istat.it/notizia/matrici-di-contiguita-distanza-e-pendolarismo/}}. 

The number of active AVCs is not constant across time; in particular, the number of municipalities hosting an AVC rose from 21 to 25 between 2021 and 2024. As of November 2024, the median road travel time to the nearest AVC was 20 minutes, with half of municipalities lying between 13 and 29 minutes and 59 municipalities (23\%) located more than 30 minutes away. Travel-time distributions were broadly similar across the four years; in Table \ref{tab:AVC_dist} we display some descriptive summaries of the road travel time for the years in scope.

\begin{table}[ht]
\begin{tabular}{llllllr}
\hline
Year & Q1      & Median  & Mean    & Q3  & Maximum & $>30$ min \\
\hline
2021 & 14'     & 20'30'' & 22'39'' & 31' & 91'     & 67 (26.2\%) \\
2022 & 12'45'' & 19'     & 21'21'' & 28' & 91'     & 58 (22.7\%) \\
2023 & 13'     & 20'     & 21'38'' & 29' & 74'     & 60 (23.4\%) \\
2024 & 13'     & 20'     & 21'18'' & 29' & 74'     & 59 (23.0\%) \\
\hline
\end{tabular}
\caption{Descriptive statistics of road travel time in minutes from each municipality to the nearest
municipality hosting an AVC, by year. The last column reports the number of
municipalities whose travel time exceeds 30 minutes.}
\label{tab:AVC_dist}
\end{table}

Considering the close relationship between socioeconomic vulnerability and both the occurrence and reporting of gender-based violence, investigated in depth by \cite{bettio2017violence}, an additional set of explanatory variables has been selected among the components of the Municipality Fragility Index (MFI) referring to year 2021 (source ISTAT$^1$). These variables capture multiple dimensions of local vulnerability, such as economic deprivation, demographic imbalance, and social marginalisation. Including them in the analysis allows us to account for the contextual conditions that may influence both the risk of experiencing violence and the likelihood of seeking help. For instance, areas characterised by higher unemployment, ageing populations, or weaker social support networks may exhibit lower reporting rates or different help-seeking patterns. These aspects will be discussed in more detail with the results of the proposed model estimation. However, among the available indicators, we excluded the distance of each municipality from the closest infrastructural pole due to the high correlation with the distance from AVCs (71\%), which in the scope of this analysis appears to be a more informative variable. This high correlation is intuitive, as AVCs are typically located in larger municipalities corresponding to infrastructure hubs.
The set of explanatory variables is summarised in Table \ref{tab:covars}. Lastly, some maps of explanatory variable are displayed in Appendix \ref{Appendix:X_maps}.

\begin{table}[!ht]
\caption{Relevant explanatory variables}
\resizebox{\textwidth}{!}{
\begin{tabular}{l l p{9cm }} 
\hline
Variable & Unit & Description\\
\hline
AVC\_dist & Minutes& Road travel time, measured in minutes, between the centroid of each municipality and the centroid of the closest municipality hosting an AVC\\
Desk\_dist & Minutes & Road travel time to the closest municipality hosting a help desk\\
ELI & Ventile & Share of employees in low productivity units in the manufacturing and services sectors. Labour productivity is given by the ratio of value added per employee and is considered low if smaller than its first quartile. \\
PGR & Rate & Population growth rate in years 2011--2021, i.e. total net migration in 2011--2021 over resident population at the end of 2011. \\
UIS & Ventile & Density of economic production units, i.e. the ratio of the stock of active business companies to the total population \\
ELL & Percentage & Percentage of people aged 25--64 with low education level, i.e. no more than middle school or vocational schools\\
PDI & Percentage & Structural dependency index, i.e. population aged $[0-19]$ or $>64$ over population aged 20--64\\
ER & Percentage & Employment rate among population aged 20--64\\
\hline
\end{tabular}
}
\label{tab:covars}
\end{table}

\section{Model implementation}\label{sec:imple}
As mentioned in the previous section, the counts of women residing in each Apulian municipality accessing AVCs within a given year $t \in \lbrace 2021, \dots, 2024 \rbrace$ are modelled via the Poisson ecological regression model described in section \ref{section:models}. In this section, we summarise the settings of the prior distributions of model parameters and describe the practical procedure to implement the model estimation in the Bayesian INLA framework. 
 
Constant covariate effects in $\beta$ and time-dependent intercepts $\beta_{0,t}$ are assigned independent $\mathcal{N}(0, 10^3)$ priors. 
For the marginal standard deviation $\sigma$ and the temporal correlation parameter $r$, we follow the PC-prior approach outlined in section \ref{subsec:PC}. Specifically, for $\sigma$ we assume $\mathrm{Prob} \lbrace \sigma^2 \leq 1/2 \rbrace= 0.9$, implying $\mathbb{E}[\sigma] \approx 0.307$ \textit{a priori}. 
Few applications can be found in the literature for the tuning of the AR(1) autocorrelation parameter $r$; in particular, \cite{fioravanti2021spatio} and \cite{franco2019unified} employed this prior in an environmental setting, where the nature of the process under study led the authors to shift prior probability mass away from zero. Here, instead, we penalise the departure from the time-independent model while attempting not to be excessively restrictive, and assume $\mathrm{Prob} \lbrace \lvert r \rvert \leq 0.4\rbrace= 0.8$. These two prior distributions for $r$ and $\sigma$ are assumed throughout the remainder of the paper. We acknowledge, however, that in our context the posterior distributions of $\sigma$ and $r$ are scarcely sensitive to hyperparameter tuning.
For the parameters $\rho$ and $\lambda$, we tested three alternative prior specifications: the PC-prior with $2/3$ probability of not exceeding the boundary value of $0.5$, 
the more restrictive PC-prior with $0.9$ probability of not exceeding the threshold of $0.6$, and the Uniform prior in $[0,1]$.

As is frequently the case in Bayesian applications, posterior inference cannot be obtained by analytical calculations. Therefore, we need to approximate the posterior distributions of all layers of model parameters. Considering that with the model at hand, covariate effects and the spatio-temporal latent field have a joint Gaussian prior and the predictor is linear, it is particularly convenient to employ the Integrated Nested Laplace Approximation \citep[INLA,][]{INLA} for the purpose. First, the joint hyperparameter posterior is Laplace approximated. Then, a Gaussian approximation is applied to the joint posterior of $\boldsymbol{\beta}$ and $\boldsymbol{z}$ conditional on hyperparameters, and the resulting posterior mean is corrected through Variational Bayes \citep{VBINLA2023}. This methodology is implemented in a dedicated \texttt{R} package (available at \url{https://www.r-inla.org/}). All spatio-temporal models are implemented within the \texttt{rgeneric} environment \citep[][chapter 11]{INLAbook} using package version \texttt{2025.06.22-1}.

All explanatory variables have been scaled to a zero mean and unit variance to facilitate the interpretation of results. Moreover, for some explanatory variables spatial trends can be noticed, as well as some major changes in the posterior distribution of the effects of two covariates in particular, namely ELI (share of low productivity workers) and ELL (low education incidence), as shown in the forest plot in Figure \ref{fig:forestplot} and discussed in more detail in the next section. This suggests the presence of spatial confounding \citep{RHZ} between these covariates and the unmeasured spatial effects. Therefore, following the simplified Spatial+ method developed by \cite{Urdangarin24} and summarised in section \ref{subsection:spatial+}, we attempt to mitigate spatial confounding by filtering out these variables' spatial trends represented by the smoothest eigenvectors of the Laplacian matrix, i.e., those associated with the smallest eigenvalues. 

The three spatial structures for the spatiotemporal Poisson model presented in section \ref{section:models}, namely ICAR, PCAR, and LCAR, are compared using the Watanabe-Akaike information criterion \citep[WAIC hereinafter,][]{GelmanWAIC}, computed internally in \texttt{R-INLA}. We also compare the three prior distributions for  $\rho$ and $\lambda$ mentioned earlier. Lastly, to evaluate the impact of deconfounding the covariates ELI and ELL, we compare, for each of the aforementioned models, the ``base'' formulation (covariates not deconfounded), and three alternative eigenvector removal patterns for ELI and ELL, i.e. the last 15, 20 or 25 Laplacian eigenvectors.

\section{Results} \label{sec:results}

Table \ref{tab:waic} reports the WAIC and its complexity component, i.e., the number of free parameters (P. eff) for all the modelling configurations described in the previous section. Differences in model performance do not appear pronounced, yet we are led to prefer the LCAR model. The treatment of spatial confounding likewise does not radically change model performance, although a small improvement can be noticed when the 20 smoothest Laplacian eigenvectors are filtered out.  
\begin{table}[ht]
\centering
\resizebox{\textwidth}{!}{
\begin{tabular}{llrrrrrrrr}
  \hline
  \multirow{2}{*}{Models} & \multirow{2}{*}{Prior} & \multicolumn{2}{c}{Base} & \multicolumn{2}{c}{S+ 15} & \multicolumn{2}{c}{S+ 20} & \multicolumn{2}{c}{S+ 25}  \\
   & &WAIC & P. eff &  WAIC & P. eff & WAIC & P. eff & WAIC & P. eff \\
  \hline
  NOSP && 4803.84 & 49.74 &  &  &  &  &  &  \\ 
  ICAR& & 3928.35 & 255.65 & 3928.15 & 255.58 & 3927.85 & 255.42 & 3928.54 & 255.88 \\ 
  PCAR&Unif & 3915.55 & 261.17 & 3914.08 & 261.02 & 3913.65 & 260.89 & 3914.69 & 261.33 \\ 
  PCAR& PC1 & 3915.11 & 261.54 & 3913.43 & 261.40 & 3913.12 & 261.23 & 3914.10 & 261.71 \\ 
  PCAR& PC2 & 3914.24 & 262.44 & 3912.51 & 262.19 & 3912.14 & 262.02 & 3913.15 & 262.52 \\ 
  LCAR&Unif & 3910.67 & 261.08 & 3908.45 & 260.83 & 3908.02 & 260.69 & 3909.38 & 261.13 \\ 
  LCAR& PC1 & 3910.27 & 261.56 & 3907.79 & 261.29 & 3907.39 & 261.15 & 3908.88 & 261.56 \\ 
  LCAR& PC2 & 3910.04 & 261.75 & 3907.47 & 261.47 & \textbf{3907.09} & 261.32 & 3908.52 & 261.75 \\
   \hline
\end{tabular}
}
\caption{Model diagnostics using the Watanabe-Akaike criterion for three spatial models with three possible Spatial+ adjustments, consisting of removing either 15, 20 or 25 Laplacian eigenvectors from the covariates ELI and ELL. For the parameter controlling the spatial association in the PCAR and LCAR we compare the Uniform prior and the PC-prior with left-tail probability values of either $2/3$ or $0.9$, associated with upper boundaries of $0.5$ or $0.6$, respectively. The P. eff column consists of the number of free parameters, i.e., the complexity component.}
\label{tab:waic}
\end{table}

In the following, we show the main results of the LCAR model after removing the 20 lowest-frequency Laplacian eigenvectors applying the simplified Spatial+ correction proposed by \cite{Urdangarin24} for two covariates: the share of employees in low-productivity firms (ELI) and the low education incidence (ELL). 

\begin{figure}[ht]
\centering
\includegraphics[width=0.9\textwidth]{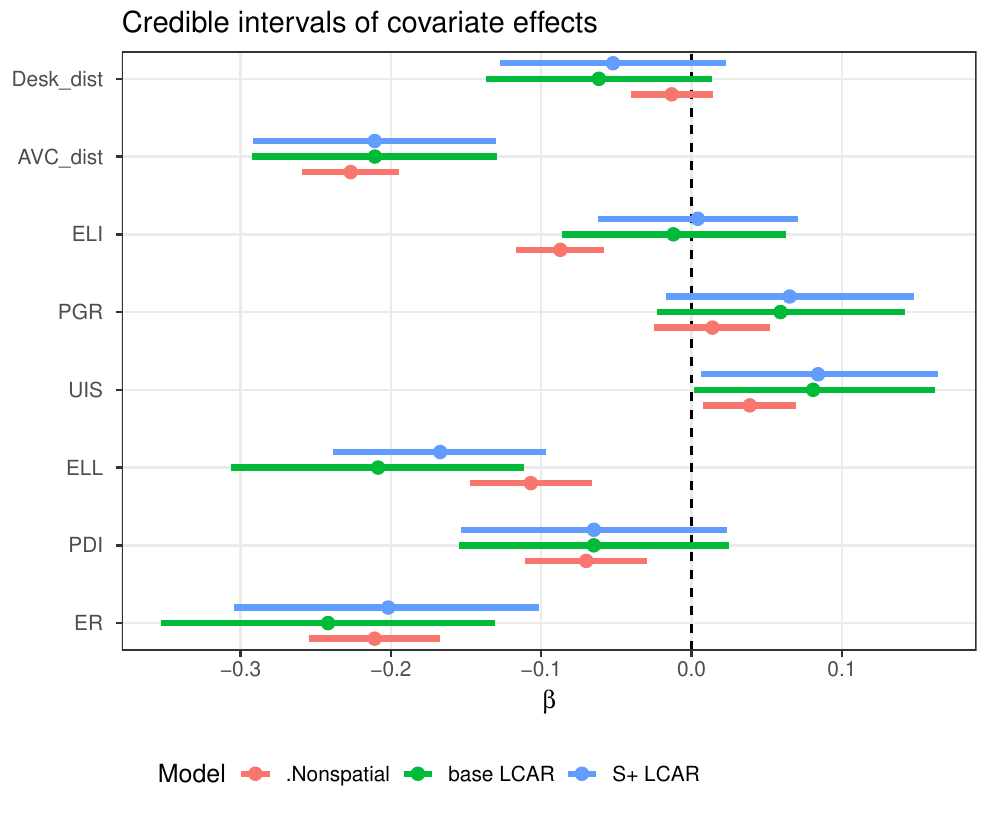}
\caption{Forest-plot of the posterior credible intervals of covariate effects using either the nonspatial model or LCAR, before ("base" models) and after filtering out the 20 lowest-frequency Laplacian eigenvectors from covariates ELI and ELL ("S+" models); LCAR hyperpriors are tuned according to Equation \ref{eq:priors} .}\label{fig:forestplot}	
\end{figure}

Hyperparameters $\sigma$, $r$, and $\lambda$ are assigned the following penalised complexity priors, as discussed in section \ref{sec:imple}:
\begin{equation}
\begin{array}{ll}
\mathrm{Prob} \lbrace \sigma^2 \leq 1/2\rbrace = 0.9 \\
\mathrm{Prob} \lbrace \lvert r \rvert \leq 0.4 \rbrace = 0.8 \\
\mathrm{Prob} \lbrace \lambda \leq 0.6\rbrace = 0.9 \\
\end{array} 
\label{eq:priors}
\end{equation}
In Figure \ref{fig:forestplot} we show a forest plot of the 95\% posterior credible intervals of covariate effects using the nonspatial and LCAR models before and after removing the 20 lowest-frequency Laplacian eigenvectors for the covariates ELI and ELL. Such a correction is motivated by the difference in the posterior distributions of these two variables between the nonspatial and spatial models, which suggests the potential confounding with the latent spatial field. In particular, including the spatial field shrinks the estimated effect of ELI towards zero, while it increases the absolute value of the estimated effect of ELL. For all covariates, the inclusion of a spatial latent field increases uncertainty in the estimation of regression coefficients, as can be seen from the wider credible intervals. The most noticeable differences among spatial models can be observed for ELL and ER; specifically, in both cases, controlling for the spatial confounding in ELI and ELL slightly reduces the covariate effect. 
In Table \ref{tab:betapost} we show in more detail the posterior summaries for intercepts and covariate effects using the LCAR model and filtering out 20 eigenvectors from ELI and ELL. 
\begin{table}[ht]
\centering
\begin{tabular}{lrrrr} 
  \hline

Effect & mean & sd & Q0.025 & Q0.975 \\ 
  \hline
  Year 2021 & -7.458 & 0.045 & -7.548 & -7.370 \\ 
  Year 2022 & -7.553 & 0.048 & -7.650 & -7.460 \\  
  Year 2023 & -7.238 & 0.041 & -7.319 & -7.158 \\ 
  Year 2024 &  -7.327 & 0.044 & -7.415 & -7.241 \\  
  AVC\_dist & -0.211 & 0.041 & -0.292 & -0.130 \\ 
  Desk\_dist & -0.052 & 0.038 & -0.128 & 0.023 \\  
  ELI & 0.004 & 0.034 & -0.062 & 0.071 \\ 
  PGR & 0.065 & 0.042 & -0.017 & 0.148 \\ 
  UIS & 0.084 & 0.040 & 0.006 & 0.164 \\ 
  ELL & -0.167 & 0.036 & -0.239 & -0.097 \\ 
  PDI & -0.065 & 0.045 & -0.153 & 0.024 \\ 
  ER & -0.202 & 0.052 & -0.305 & -0.102 \\ 
   \hline
\end{tabular}
\captionsetup{width=0.8\textwidth}
\caption{Posterior summaries of year-specific intercepts and covariate effects when the spatial component is modelled as a LCAR process with hyperpriors defined as in Equation \ref{eq:priors}. Simplified spatial+ correction has been applied to ELI and ELL covariates by removing the 20 lowest-frequency Laplacian eigenvectors.}
\label{tab:betapost}
\end{table}
To assess the effects of explanatory variables on AVC accesses, we need to recall the twofold interpretation of the response variable, as it depends on both the frequency at which violence incidents occur and the frequency at which victims seek help when it occurs.
The three strongest predictors across the years in scope appear to be the travel time from the closest municipality hosting an AVC (AVC\_dist), the percentage of people with a low educational level (ELL), and the employment rate (ER), in all cases with a negative association.
Additionally, the density of productive units (UIS) has a positive association, and the credible interval of its coefficient ranges entirely away from zero. 

The negative association between AVC accesses and distance from AVCs can be read as a symptom of the persistent physical barrier to seeking help. The magnitude of the coefficient may be interpreted as follows: an increase in the travel time to the closest AVC by one standard deviation, i.e. approximately $13'40''$, is associated, \textit{ceteris paribus}, with a decrease of about $18.9\%$ in the frequency of accesses to AVCs, since $\exp\lbrace E\left[\beta_\mathrm{AVC\_Dist}|y \right]\rbrace = 0.811$. The expected association with the distance from the closest help desk (Desk dist) is negative, consistent with the distance from the closest AVC, but the magnitude of such expected association is smaller, and, most importantly, the credible interval includes zero. This result may suggest that distance to help desks poses a weaker barrier to seeking assistance than distance to AVCs, possibly reflecting the wider geographical distribution of help desks.

In light of the twofold process generating AVC accesses, a tentative yet intuitive interpretation could be that the negative association with low education may operate through a reduced propensity to seek help, while the negative association with the employment rate may reflect a lower occurrence of violence in more economically developed contexts. This latter consideration should be interpreted taking into consideration the state-of-the-art evidence that both women's and partners' unemployment are associated with higher violence occurrence, as reviewed by \cite{bettio2017violence}. The positive association of accesses with the density of productive units (UIS), instead, is less intuitive, as a higher number of firms does not imply \textit{per se} higher economic development, even though it has a negative weight in determining the Municipality Fragility Index.

 Lastly, hyperparameter posterior summaries for the LCAR model after ELI and ELL have been corrected for spatial confounding are shown in Table \ref{tab:hyperpar}.
\begin{table}[ht]
\centering
\begin{tabular}{lrrrrr}
  \hline
  param & mean & sd & Q0.025 & Median & Q0.975 \\ 
  \hline
 $\lambda$ & 0.558 & 0.155 & 0.253 & 0.562 & 0.836 \\  
  $\sigma$ & 0.647 & 0.060 & 0.537 & 0.644 & 0.772 \\ 
  $r$ & 0.600 & 0.061 & 0.471 & 0.603 & 0.711 \\
   \hline
\end{tabular}
\captionsetup{width=0.8\textwidth}
\caption{Posterior summaries of the hyperparameters of the AR(1) -- LCAR latent spatiotemporal effect with hyperpriors defined as in Equation \ref{eq:priors}.}
\label{tab:hyperpar}
\end{table}
Spatial dependence does not appear particularly strong, as $\lambda$ has a posterior expectation of $\approx 0.56$, which means that neighbouring sites display a moderate degree of conditional association. The standard deviation of the spatio-temporal latent field has a posterior expectation of $\approx 0.65$. This can be interpreted such that an increment of 1 standard deviation of the latent field in a municipality implies an increase of approximately $\exp(0.647)-1\approx 91\%$ in the frequency of AVC accesses, although the value of $\sigma$ cannot be disentangled from the effect induced by the adjacency structure, as proper CAR models are not scalable like intrinsic models \citep{Sorbye}; therefore, this interpretation is only valid with regard to the specific territorial domain herein considered.
Lastly, the temporal autocorrelation is positive, as one would expect by comparing the spatial patterns in AVC accesses in Figure \ref{fig:log_access}.


\section{Discussion and concluding remarks}

\label{section:discussion}

In this work, the analysis of help-seeking behaviour of women who suffered gender-based violence provided the motivation for an original Bayesian spatio-temporal modelling framework and for methodological contributions focused on principled prior construction.

A methodological choice in this study was to restrict the analysis to CAR-type models. Although the BYM model could potentially improve predictive accuracy, its inclusion would shift the inferential focus from the strength of spatial association to the allocation of residual variance between spatial structure and random noise, making the comparison less aligned with the aims of this study.
A different and arguably heavier limitation is that we must presently employ a single predictor combining the two processes of violence occurring and help seeking after violence occurs, which cannot be disentangled using available data. Consequently, the effects we estimate reflect both mechanisms at once. Additional limitations include the lack of individual-level information for privacy reasons 
and the usual constraints posed by aggregated data that can mask within-area heterogeneity. 

A promising development would be the implementation of a Poisson-logistic regression model \citep{Stoner02102019}, in line with the analysis of \cite{Polettini}, to allow the explicit quantification of VAW under-reporting. To this end, however, more integrated datasets are required combining administrative, health, social-service, and AVC-level information, thereby enabling the identification of the respective contributions of structural incidence, help-seeking propensity and territorial accessibility, and allowing the two processes of occurrence and help-seeking to be modelled separately. 

It is worth discussing which features of the proposed modelling framework can be transferred directly to other applications and which should be elicited case by case. The PC-priors defined in sections \ref{subsection:PC_rho} and \ref{subsection:PC_lambda} depend on the areal domain through the spectra of deterministic matrices and are independent of any other parameter, while the PC-prior for the temporal autocorrelation only depends on the number of time points. Therefore, no step of the derivation of these priors needs to be repeated when the same models are applied to different areal domains -- a natural application instance would be, e.g., disease mapping.

What should instead be elicited case by case are the probabilistic assumptions encoded in the PC-prior of each parameter, namely the boundary value and its associated tail or central probability. This is especially relevant for the standard deviation $\sigma$: since proper CAR models are not scalable in the sense of \cite{Sorbye}, the interpretation of $\sigma$ depends on the adjacency structure of the specific spatial domain, so that its boundary value cannot be transferred automatically across different applications.
 
Overall, the present results show that seeking support is far from a straightforward process: the negative association between access rates and proximity to the nearest AVC suggests that even modest distances can act as barriers to help-seeking. Moreover, education-related vulnerability and local economic conditions appear to affect the likelihood of accessing AVCs, suggesting that structural factors, cultural attitudes, and service accessibility interact in shaping women’s behaviour. Even after adjusting for potential spatial confounding, a residual spatial effect remains, pointing to unobserved factors that may influence women’s ability to activate support pathways. Beyond the empirical findings, the methodological contribution of this work lies in the derivation of PC-priors for the PCAR and LCAR spatial association parameters, which may contribute to improving interpretability and stabilising INLA-based inference. However, extending PC-priors for areal model hyperparameters where not already implemented in the literature, e.g. for more flexible multivariate models not covered in this analysis, may constitute a stand-alone line of forthcoming research, to be supported by a broader analysis and simulation studies.

\section*{Declarations}
\begin{small}
\subsubsection*{Code and data availability}
The source R code is available at this address: \url{https://github.com/lcef97/CAV_Puglia/blob/main/CAV_output.R}. Exact replicability of results depends on  the \texttt{R-INLA} version used; authors used the testing version \texttt{2025.06.22-1}.
\subsubsection*{Acknowledgements}
Leonardo Cefalo wishes to thank Professor H\r{a}vard Rue from the King Abdullah University of Science and Technology for his valuable assistance in troubleshooting and initialising the INLA optimisation algorithm. 
\subsubsection*{Funding}
This study was partially funded by the European Union -- NextGenerationEU, Mission 4, Component 2, in the framework of the GRINS -- Growing Resilient, INclusive and Sustainable project (GRINS PE00000018 -- CUP H93C22000650001 and CUP J33C22002910001). 
\\
The study was also partially funded by the European Union -- NextGenerationEU, Mission 4, Component 1, in the framework of the PRIN project ``Violence against women: modelling misreported information in social data'' (CUP H53D23006060006). 
This project benefits from collaborating with the Statistical Section and the Welfare Department of the Apulia Region (Italy), which actively monitors gender-based violence and conducts an annual survey to analyse its trends and impacts. Their ongoing efforts in data collection and research have been essential to the development of this study.
\\
The views and opinions expressed are solely those of the authors and do not necessarily reflect those of the European Union, nor can the European Union be held responsible for them.

\end{small}

\bibliographystyle{apalike}

\bibliography{References}

\begin{appendices}

\section{PC-prior derivation details} \label{Appendix:PCpriors}

In the case of latent Gaussian models, the KLD between any pair of distributions is given by a simple formula, as proved by \cite{PC}. Specifically, the KLD between two zero-mean $Tn$--variate Gaussian distributions differing in their variance-covariance matrices, say $\boldsymbol{\mathrm{Var}_0}$ and $\boldsymbol{\mathrm{Var}_1}$ for the base and flexible formulation respectively, reads as \citep{PC}:

\begin{equation}\label{eq:KLD}
\mathrm{KLD(\mathcal{M}_1\lvert \rvert\mathcal{M}_0)} = -\frac{1}{2} \ln
\begin{vmatrix}\boldsymbol{\mathrm{Var}_0}^{-1} \boldsymbol{\mathrm{Var}_1}\end{vmatrix} 
- \frac{Tn}{2} + \frac{1}{2} \mathrm{tr}\left( \boldsymbol{\mathrm{Var}_0}^{-1} \boldsymbol{\mathrm{Var}_1} \right)
\end{equation}

Given the KLD expression in Equation (\ref{eq:KLD}), the prior can be derived with a change of variable from the KLD to the parameter through which $\mathcal{M}_1$ and $\mathcal{M}_0$ differ. With no loss of generality, for a generic parameter $\theta$, which is random in $\mathcal{M}_1$ and fixed in $\mathcal{M}_0$, if the KLD is monotonic, the prior reads as
\begin{equation}
\pi(\theta) = \psi_\theta e^{
-\psi_\theta\sqrt{
2\mathrm{KLD}(\mathcal{M}_1\lvert \rvert\mathcal{M}_0)}}
\frac{1}{\sqrt{2\mathrm{KLD}(\mathcal{M}_1\lvert \rvert\mathcal{M}_0)}}
\left|\frac{\partial\mathrm{KLD}(\mathcal{M}_1\lvert \rvert\mathcal{M}_0)}{\partial\theta} \right|
\label{eq:PC_general}
\end{equation}
for a rate parameter $\psi_\theta$ which is in turn controlled through a boundary value for $\theta$ and an associated tail probability. In the following, we provide some details on the results outlined in section \ref{subsec:PC}. 
If instead the KLD is an even function, as for the AR(1) temporal autocorrelation, the right-hand side of the Equation above is to be divided by 2.

\subsection{AR(1) autocorrelation in spatio-temporal models} \label{PC_AR1_detail}

In this case, we are comparing the flexible AR(1) temporal structure $\boldsymbol{\mathrm{Var}_1} = \frac{\sigma^2}{1-r^2}\boldsymbol{R} \otimes\boldsymbol{\Omega}$, against the base model in which no temporal autocorrelation is allowed, i.e. $\boldsymbol{\mathrm{Var}_0} = \sigma^2\boldsymbol{I_T}\otimes\boldsymbol{\Omega}$, where $\boldsymbol{R}$, $\boldsymbol{\Omega}$, $r$ and $\sigma^2$ have been defined as in section \ref{section:models}. Considering Equation \ref{eq:KLD}, the KLD for $r$ reads as:
  \begin{equation}
\mathrm{KLD}\left(\pi\left(\boldsymbol{z} \vert r\right)\, || \,\pi\left(\boldsymbol{z} \vert r=0\right)\right) =   \frac{n}{2}
 \left( \ln(1-r^2)
 +T  \frac{r^2}{1-r^2} \right)
 \label{eq:KLD_AR1}
 \end{equation}
This is a U-shaped even function of $r$, with a minimum at 0 for $r=0$ and tending to infinity for $\lvert r \rvert$ approaching 1. Although the KLD depends on the number of sites $n$, the PC-prior can be seen to be independent of it. To see this, consider from Equation \ref{eq:KLD_AR1} that $ \mathrm{KLD}\left(\pi\left(\boldsymbol{z} \vert r\right)\, || \,\pi \left(\boldsymbol{z} \vert r=0\right)\right) = \frac{n}{2}h(r)$ where $h(r)$ does not depend on $n$; then, assuming an $\mathrm{exp}(\psi_r)$ distribution on $\sqrt{2\mathrm{KLD}\left(\pi\left(\boldsymbol{z} \vert r\right)\, || \,\pi\left(\boldsymbol{z} \vert r=0\right)\right)}$ and parametrising the PC-prior through the central probability $\alpha$ and the upper boundary $U$, we have 
 $$\psi_r = \ln \frac{1}{1-\alpha} \frac{1}{\sqrt{n h(U)}}$$

Given the KLD formula, if we apply the change of variable as in Equation \ref{eq:PC_general}, and express $\psi_r$ in terms of $\alpha$ and $U$, we obtain Equation \ref{eq:PC_AR1}. 

\subsection{Spatial association parameters}

In this case, we are comparing two models that differ through the spatial covariance matrix $\boldsymbol{\Omega}$, which in turn depends on a spatial association parameter. With no loss of generality, we thus have for the flexible case $\boldsymbol{\mathrm{Var}_1} = \frac{\sigma^2}{1-r^2}\boldsymbol{R} \otimes \boldsymbol{\Omega}$, where $\boldsymbol{\Omega}$ is defined, depending on the specific model in scope, as in section \ref{subsection:spatial}, and $\boldsymbol{\mathrm{Var}_0} = \frac{\sigma^2}{1-r^2}\boldsymbol{R} \otimes\boldsymbol{\Omega_0}$, where $\boldsymbol{\Omega_0} = \boldsymbol{I_n}$ in the LCAR and $\boldsymbol{\Omega_0} = \boldsymbol{D}^{-1}$ in the PCAR, $\boldsymbol{D}$ being the graph degree matrix, which in the following we assume to be invertible. Recalling Equation \ref{eq:KLD}, it is sufficient to compute the eigenvalues of $\boldsymbol{\mathrm{Var}_0}^{-1} \boldsymbol{\mathrm{Var}_1} = \boldsymbol{I_T}\otimes \left( \boldsymbol{\Omega_0}^{-1} \boldsymbol{\Omega}\right)$, which depend only on the model-specific spatial association parameter. In the following, we provide some details on how the KLD is obtained for the PCAR and LCAR models.

\subsubsection{PCAR model} \label{appendix_pcar}

For this model, we have $\boldsymbol{\Omega_0}^{-1} \boldsymbol{\Omega} = \boldsymbol{D}\left( \boldsymbol{D}-\rho \boldsymbol{W} \right)^{-1}$, whose eigenvalues are given by $\frac{1}{1-\rho \delta_i} \forall i \in \lbrace1, \dots, n\rbrace$, where $\delta_i$ is the generic $i$-th eigenvalue of the deterministic matrix $\boldsymbol{D^{-1}W}$. Considering that this matrix is row-stochastic with null diagonal, due to the Gershgorin circle theorem \citep{HornJohnson}, we have $\lvert\delta_i \rvert \leq 1$ $\forall i$. Specifically, the highest element of $\delta$ is equal to 1 and is associated with a constant eigenvector. Moreover, since $\boldsymbol{D^{-1}W}$ is similar to $\boldsymbol{D^{-1/2} W D^{-1/2}}$, all the elements of $\delta$ are real.  
Then, we have
$$
\ln\left| \boldsymbol{\mathrm{Var}_0}^{-1} \boldsymbol{\mathrm{Var}_1}\right| = -T \sum_{i=1}^n \ln\left(1-\rho\delta_i\right)
$$
and
$$
\mathrm{tr}\left( \boldsymbol{\mathrm{Var}_0}^{-1} \boldsymbol{\mathrm{Var}_1}\right) = T \sum_{i=1}^n \frac{1}{1-\rho \delta_i}
$$
Substituting these terms in Equation \ref{eq:KLD}, we obtain Equation \ref{eq:KLD_PCAR}. Taking the derivative, we obtain
$$
\frac{\partial \mathrm{KLD}
\left(\pi\left(\boldsymbol{z} \vert\rho \right)\, || \,
\pi \left(\boldsymbol{z} \vert\rho=0\right)\right)}
{\partial \rho} = 
\frac{T\rho}{2} \sum_{i=1}^n \frac{\delta_i^2}{(1-\rho\delta_i)^2}
$$
which is a positive quantity for $\rho \in(0,1)$ and equals zero at $\rho=0$. Then we can see that the KLD is a monotonically increasing function, with limit for $\rho \rightarrow1 $ given by $\infty$.

\subsubsection{LCAR model} \label{appendix_lcar}

For this model, we have $\boldsymbol{\Omega_0}^{-1} \boldsymbol{\Omega} = \left(\lambda\boldsymbol{L} + \left(1-\lambda\right)\boldsymbol{I_n}\right)^{-1}$, whose eigenvalues are given by $\frac{1}{\lambda (\ell_i -1) + 1 } \forall i \in \lbrace1, \dots, n\rbrace$, where $\ell_i$ is the generic $i$-th eigenvalue of the graph Laplacian $\boldsymbol{L}$. For the determinant and trace terms, we have
$$
\ln\left| \boldsymbol{\mathrm{Var}_0}^{-1} \boldsymbol{\mathrm{Var}_1}\right| = -T \sum_{i=1}^n \ln\left(\lambda(\ell_i - 1) + 1\right)
$$
and
$$
\mathrm{tr}\left( \boldsymbol{\mathrm{Var}_0}^{-1} \boldsymbol{\mathrm{Var}_1}\right) = T \sum_{i=1}^n \frac{1}{\lambda(\ell_i - 1) + 1}
$$
Since all Laplacian eigenvalues are nonnegative, $\lambda \in [0,1)$ is a sufficient condition for $\lambda \left(\ell_i - 1\right)+1$ to be strictly positive for each $i$. Moreover, considering that at least one Laplacian eigenvalue is zero, it can be seen from Equation \ref{eq:KLD_LCAR} that the KLD tends to infinity as $\lambda$ approaches one. Substituting the determinant and trace formulas above into Equation \ref{eq:KLD}, we obtain Equation \ref{eq:KLD_LCAR}.
Taking the derivative of the KLD, we obtain
$$
\frac{\partial \mathrm{KLD}
\left(\pi\left(\boldsymbol{z} \vert\lambda)\, || \,
\pi(\boldsymbol{z} \vert\lambda=0\right)\right)}
{\partial \lambda} = \frac{T\lambda}{2} \sum_{i=1}^n
\frac{ (\ell_i - 1)^2}{(1+\lambda (\ell_i - 1) )^2}
$$
which is, again, a positive quantity for $\lambda \in (0, 1)$ and equals zero at $\lambda=0$. As for the PCAR, the KLD is a monotonically increasing function.

\section{Maps of explanatory variables} \label{Appendix:X_maps}
In this section, we display some additional maps, i.e. the distance of each municipality from the closest one hosting an AVC, the components of the Municipality Fragility Index (MFI) included as explanatory variables in the analytical model described in section \ref{section:models}, and the distance from municipalities hosting a help desk. The population growth rate map does not include the municipality of Anzano di Puglia (FG) due to its extreme record ($ -311\%$). Regarding the distance from AVCs, for the sake of brevity we only show it for the first and last year, corresponding to the minimum and maximum number of active centres, respectively.

\begin{figure}

\subfloat{\includegraphics[width = 2.5in]{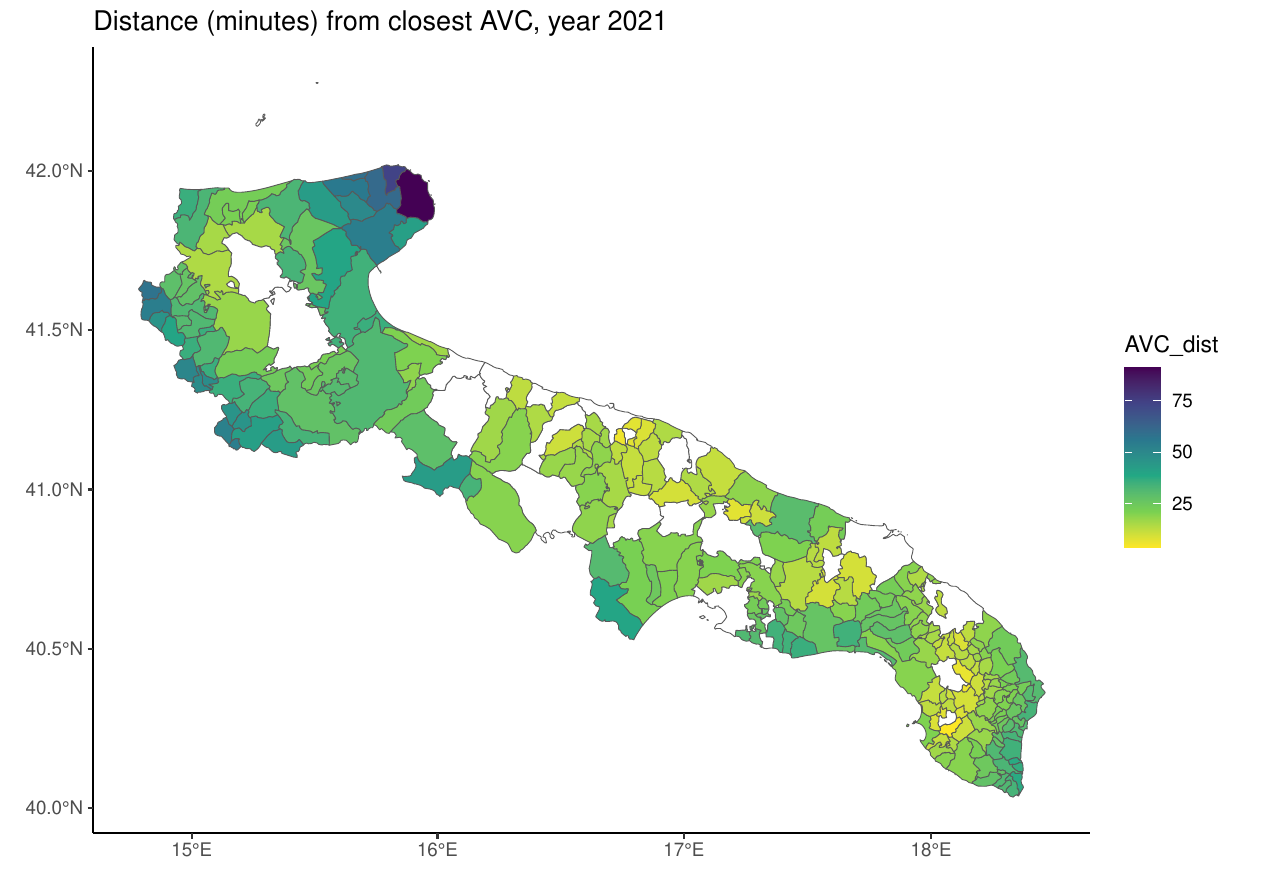}} 
\subfloat{\includegraphics[width = 2.5in]{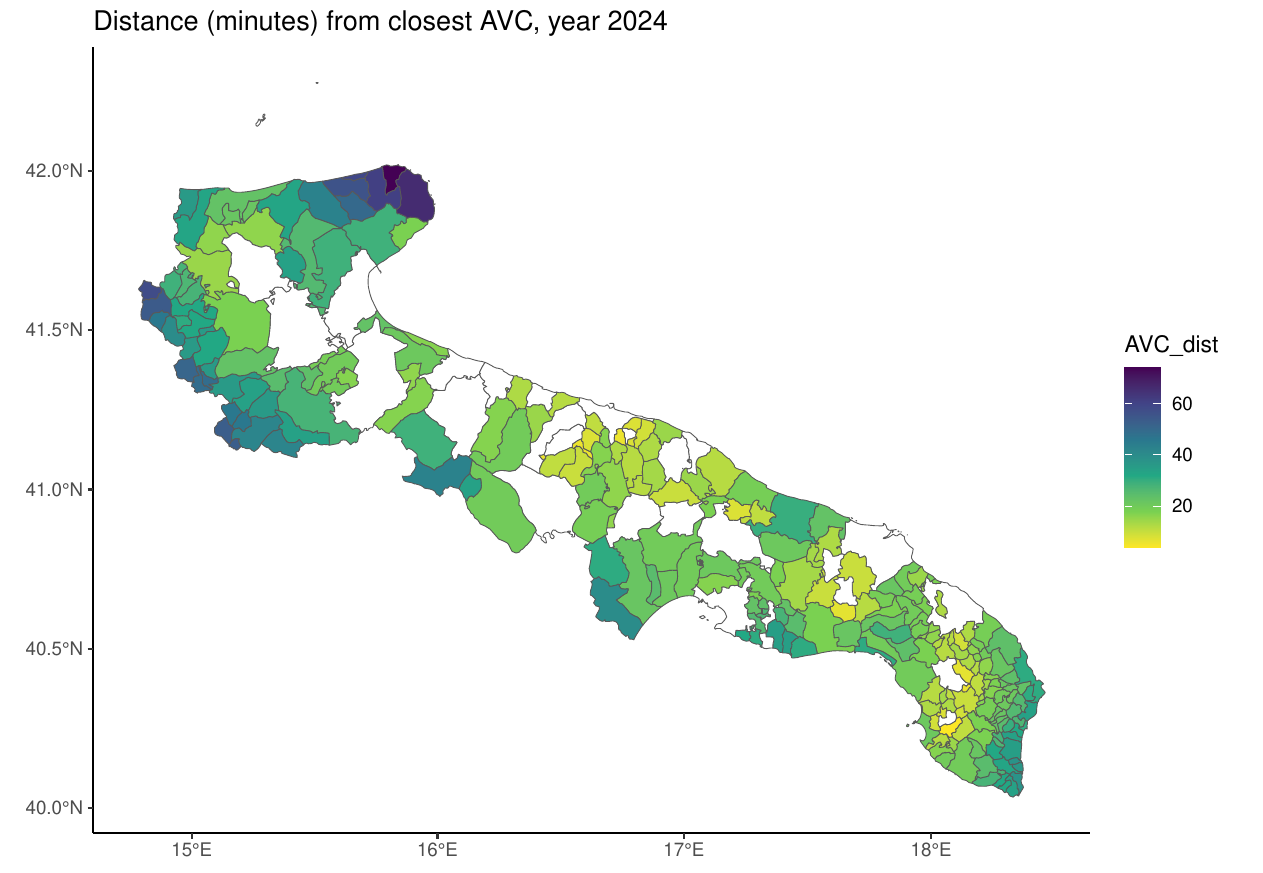}}\\
\subfloat{\includegraphics[width = 2.5in]{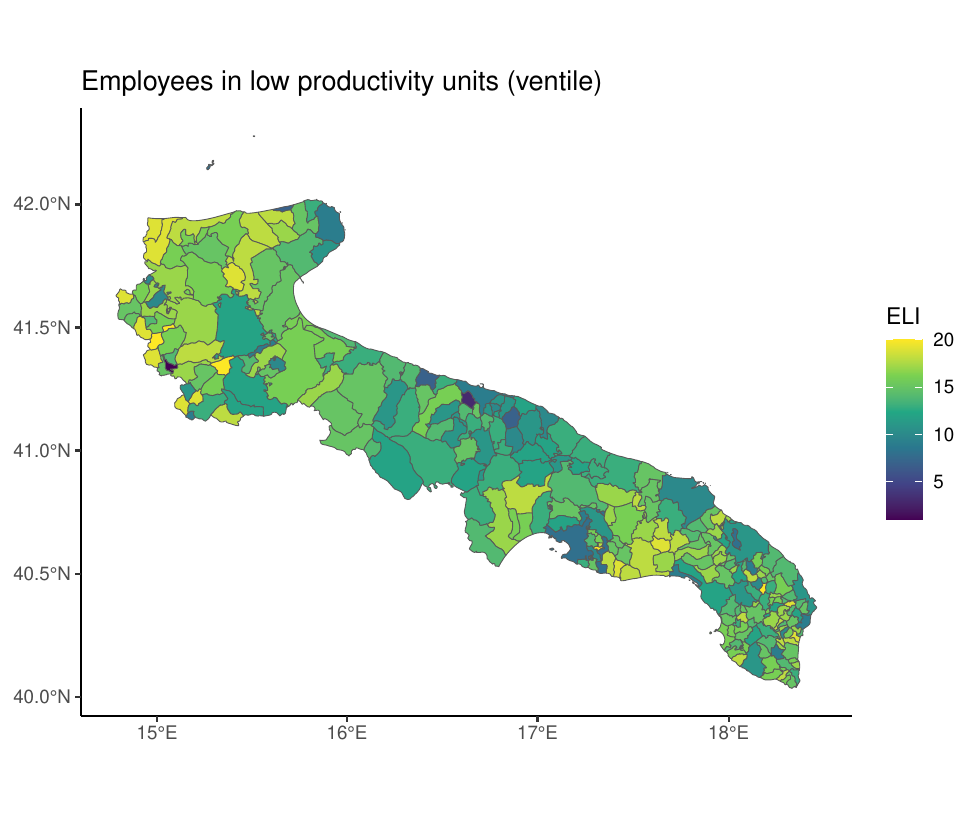}} 
\subfloat{\includegraphics[width = 2.5in]{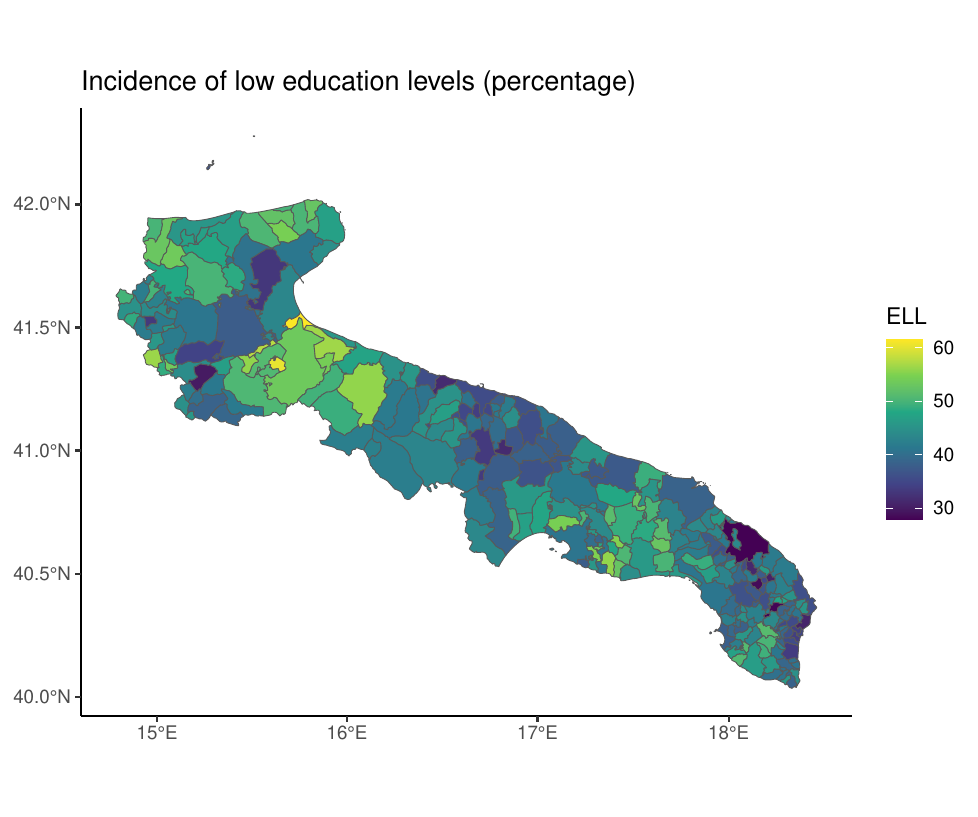}}\\
\subfloat{\includegraphics[width = 2.5in]{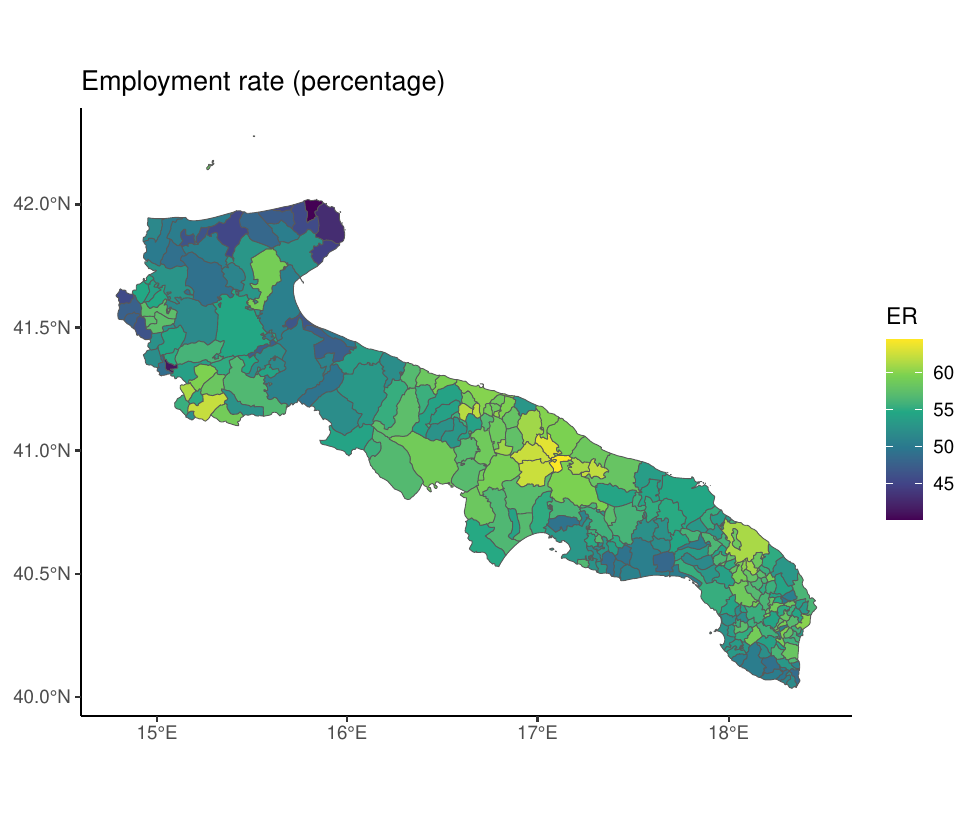}}
\subfloat{\includegraphics[width = 2.5in]{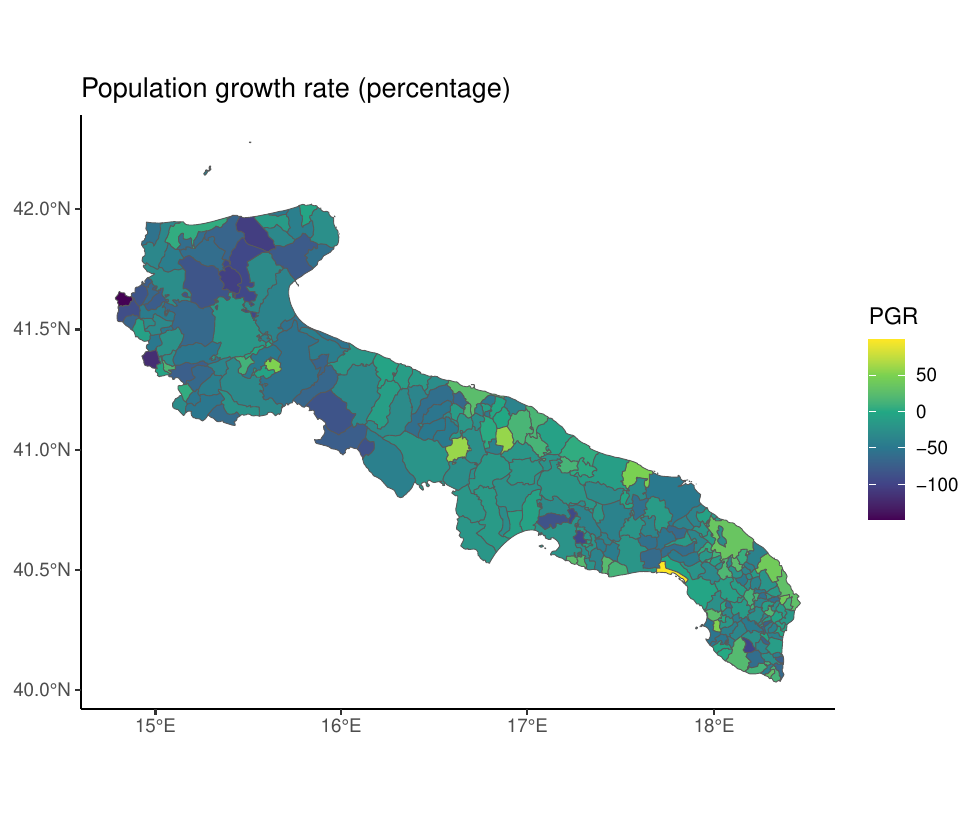}} \\
\subfloat{\includegraphics[width = 2.5in]{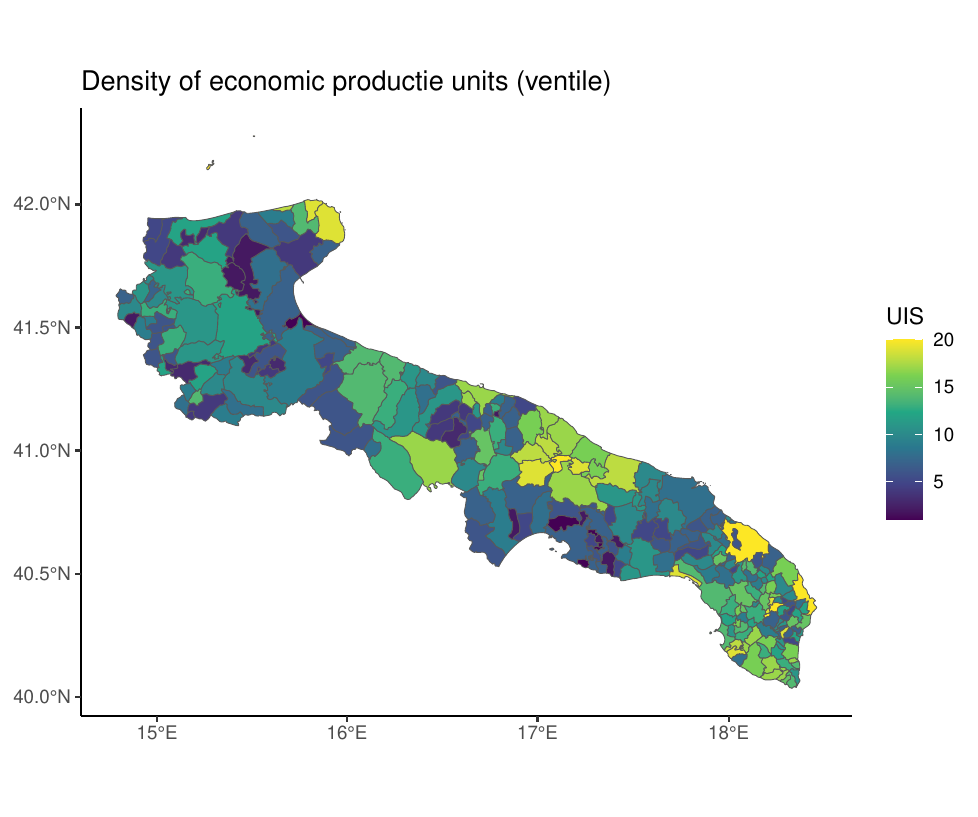}} 
\subfloat{\includegraphics[width = 2.5in]{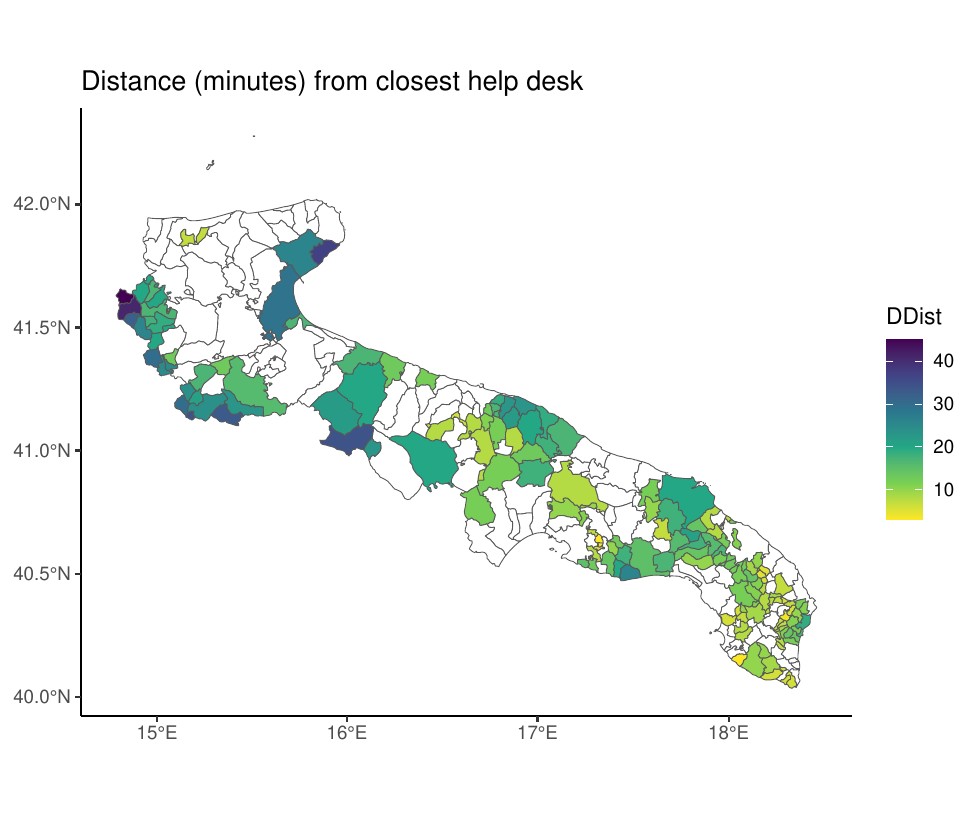}} 
\caption{Additional explanatory variables: distance from AVCs in years 2021 and 2024, MFI components and distance from help desks. White areas in the two maps on the top and in the one on the bottom right correspond to municipalities in which an AVC or a help desk is available, respectively (distance is zero).}
\label{fig:covariates}
\end{figure}

\end{appendices}

\end{document}